\documentclass[%
 reprint,
superscriptaddress,
 amsmath,amssymb,
 aps,
nofootinbib
]{revtex4-2}

\usepackage{graphicx}
\usepackage{dcolumn}
\usepackage{bm}
\usepackage{subfig}
\usepackage{float}
\usepackage{xcolor}
\usepackage{ulem}
\usepackage{hyperref}
\usepackage[output-decimal-marker={.},exponent-product=\cdot]{siunitx}

\usepackage[english]{babel}
\definecolor{darkgreen}{rgb}{0.2,0.5, 0.2}

\definecolor{mbscolor}{rgb}{0.60, 0.0, 0.65}

\usepackage{amsmath}

\begin{document}


\title{Binding energies of ground and isomeric states in neutron-rich ruthenium isotopes: measurements at JYFLTRAP and comparison to theory}

\author{M.~Hukkanen}
\affiliation{University of Jyvaskyla, Department of Physics, Accelerator Laboratory, P.O. Box 35(YFL) FI-40014 University of Jyvaskyla, Finland}
\affiliation{Universit\'e de Bordeaux, CNRS/IN2P3, LP2I Bordeaux, UMR 5797, F-33170 Gradignan, France}
\author{W.~Ryssens}
\affiliation{Institut d'Astronomie et d'Astrophysique, Universit\'e Libre de Bruxelles, Campus de la Plaine CP 226, 1050 Brussels, Belgium}
\author{P.~Ascher}
\affiliation{Universit\'e de Bordeaux, CNRS/IN2P3, LP2I Bordeaux, UMR 5797, F-33170 Gradignan, France}
\author{M.~Bender}
\affiliation{Universit\'e de Lyon, Universit\'e Claude Bernard Lyon 1, CNRS/IN2P3, IP2I Lyon, UMR 5822, F-69622 Villeurbanne, France}
\author{T.~Eronen}
\affiliation{University of Jyvaskyla, Department of Physics, Accelerator Laboratory, P.O. Box 35(YFL) FI-40014 University of Jyvaskyla, Finland}
\author{S.~Gr\'evy}
\affiliation{Universit\'e de Bordeaux, CNRS/IN2P3, LP2I Bordeaux, UMR 5797, F-33170 Gradignan, France}
\author{A.~Kankainen}
\affiliation{University of Jyvaskyla, Department of Physics, Accelerator Laboratory, P.O. Box 35(YFL) FI-40014 University of Jyvaskyla, Finland}
\author{M.~Stryjczyk}
\affiliation{University of Jyvaskyla, Department of Physics, Accelerator Laboratory, P.O. Box 35(YFL) FI-40014 University of Jyvaskyla, Finland}
\author{L.~Al~Ayoubi}
\affiliation{University of Jyvaskyla, Department of Physics, Accelerator Laboratory, P.O. Box 35(YFL) FI-40014 University of Jyvaskyla, Finland}
\affiliation{Universit\'e Paris Saclay, CNRS/IN2P3, IJCLab, 91405 Orsay, France}
\author{S.~Ayet}
\affiliation{II. Physikalisches Institut, Justus Liebig Universit\"at Gie{\ss}en, 35392 Gie{\ss}en, Germany}
\author{O.~Beliuskina}
\affiliation{University of Jyvaskyla, Department of Physics, Accelerator Laboratory, P.O. Box 35(YFL) FI-40014 University of Jyvaskyla, Finland}
\author{C.~Delafosse}
\altaffiliation[Present address: ]{Universit\'e Paris Saclay, CNRS/IN2P3, IJCLab, 91405 Orsay, France}
\affiliation{University of Jyvaskyla, Department of Physics, Accelerator Laboratory, P.O. Box 35(YFL) FI-40014 University of Jyvaskyla, Finland}
\author{Z.~Ge}
\affiliation{GSI Helmholtzzentrum f\"ur Schwerionenforschung, 64291 Darmstadt, Germany}
\author{M.~Gerbaux}
\affiliation{Universit\'e de Bordeaux, CNRS/IN2P3, LP2I Bordeaux, UMR 5797, F-33170 Gradignan, France}
\author{W.~Gins}
\affiliation{University of Jyvaskyla, Department of Physics, Accelerator Laboratory, P.O. Box 35(YFL) FI-40014 University of Jyvaskyla, Finland}
\author{A.~Husson}
\affiliation{Universit\'e de Bordeaux, CNRS/IN2P3, LP2I Bordeaux, UMR 5797, F-33170 Gradignan, France}
\author{A.~Jaries}
\affiliation{University of Jyvaskyla, Department of Physics, Accelerator Laboratory, P.O. Box 35(YFL) FI-40014 University of Jyvaskyla, Finland}
\author{S.~Kujanp\"a\"a}
\affiliation{University of Jyvaskyla, Department of Physics, Accelerator Laboratory, P.O. Box 35(YFL) FI-40014 University of Jyvaskyla, Finland}
\author{M.~Mougeot}
\affiliation{University of Jyvaskyla, Department of Physics, Accelerator Laboratory, P.O. Box 35(YFL) FI-40014 University of Jyvaskyla, Finland}
\author{D.A.~Nesterenko}
\affiliation{University of Jyvaskyla, Department of Physics, Accelerator Laboratory, P.O. Box 35(YFL) FI-40014 University of Jyvaskyla, Finland}
\author{S.~Nikas}
\affiliation{University of Jyvaskyla, Department of Physics, Accelerator Laboratory, P.O. Box 35(YFL) FI-40014 University of Jyvaskyla, Finland}
\author{H.~Penttil\"a}
\affiliation{University of Jyvaskyla, Department of Physics, Accelerator Laboratory, P.O. Box 35(YFL) FI-40014 University of Jyvaskyla, Finland}
\author{I.~Pohjalainen}
\affiliation{University of Jyvaskyla, Department of Physics, Accelerator Laboratory, P.O. Box 35(YFL) FI-40014 University of Jyvaskyla, Finland}
\author{A.~Raggio}
\affiliation{University of Jyvaskyla, Department of Physics, Accelerator Laboratory, P.O. Box 35(YFL) FI-40014 University of Jyvaskyla, Finland}
\author{M.~Reponen}
\affiliation{University of Jyvaskyla, Department of Physics, Accelerator Laboratory, P.O. Box 35(YFL) FI-40014 University of Jyvaskyla, Finland}
\author{S.~Rinta-Antila}
\affiliation{University of Jyvaskyla, Department of Physics, Accelerator Laboratory, P.O. Box 35(YFL) FI-40014 University of Jyvaskyla, Finland}
\author{A.~de Roubin}
\altaffiliation[Present address: ]{KU Leuven, Instituut voor Kern- en Stralingsfysica, B-3001 Leuven, Belgium}
\affiliation{University of Jyvaskyla, Department of Physics, Accelerator Laboratory, P.O. Box 35(YFL) FI-40014 University of Jyvaskyla, Finland}
\author{J.~Ruotsalainen}
\affiliation{University of Jyvaskyla, Department of Physics, Accelerator Laboratory, P.O. Box 35(YFL) FI-40014 University of Jyvaskyla, Finland}
\author{V.~Virtanen}
\affiliation{University of Jyvaskyla, Department of Physics, Accelerator Laboratory, P.O. Box 35(YFL) FI-40014 University of Jyvaskyla, Finland}
\author{A.P.~Weaver}
\altaffiliation[Present address: ]{TRIUMF, 4004 Wesbrook Mall, Vancouver, British Columbia V6T 2A3, Canada}
\affiliation{School of Computing, Engineering and Mathematics, University of Brighton, Brighton BN2 4GJ, United Kingdom}

\date{\today}



\begin{abstract}
We report on precision mass measurements of $^{113,115,117}$Ru performed with the JYFLTRAP double Penning trap mass spectrometer at the Accelerator Laboratory of University of Jyv\"askyl\"a. The phase-imaging ion-cyclotron-resonance technique was used to resolve the ground and isomeric states in $^{113,115}$Ru and enabled for the first time a measurement of the isomer excitation energies, $E_x(^{113}$Ru$^{m})=100.5(8)$~keV and  $E_x(^{115}$Ru$^{m})=129(5)$~keV. The ground state of $^{117}$Ru was measured using the time-of-flight ion-cyclotron-resonance technique. The new mass-excess value for $^{117}$Ru is around 36 keV lower and 7 times more precise than the previous literature value. With the more precise ground-state mass values, the evolution of the two-neutron separation energies is further constrained and a similar trend as predicted by the BSkG1 model is obtained up to the neutron number $N=71$.
\end{abstract}

\maketitle


\section{\label{sec:intro} Introduction}

Neutron-rich nuclei between zirconium ($Z=40$) and tin ($Z=50$) exhibit a variety of shapes; several of them even exhibit shape coexistence, where excited states are linked to shapes which differ from that of the nuclear ground state. 
The diverse manifestations of collectivity in general and nuclear shapes in particular in this region of the nuclear chart have been studied widely, both theoretically and experimentally, see e.g. Ref. \cite{Garrett2022} and references therein. The relevant nuclear configurations are not limited to shapes with a comparatively high degree of symmetry such as spheres or axially symmetric ellipsoids with prolate or oblate deformation, but also includes shapes with no remaining rotational symmetry axis: triaxial shapes. There is evidence that the ground states of neutron-rich ruthenium isotopes ($Z=44$) fall in the latter category~\cite{Garrett2022,Srebrny2006}, an interpretation that is further supported by different models~\cite{Moller2006,Scamps2021}. These models typically agree that the effect of triaxial deformation is largest at the mid-shell and that the effect tapers off when even more neutrons are added to the nucleus, i.e. that sufficiently neutron-rich nuclei revert to an axially symmetric or spherical shape towards the shell closure at $N=82$.


Structural changes can be studied via a wide range of experimental methods, including laser- and decay-spectroscopy as well as Coulomb excitation. At the same time, Penning-trap mass spectrometry can be used to explore differences in binding energy which can reveal possible shape transitions~\cite{Hager2006,Naimi2010,Chaudhuri2013}. With the development of the phase-imaging ion-cyclotron-resonance (PI-ICR) technique~\cite{Eliseev2013,Eliseev2014}, not only the ground-state binding energies but also the isomer excitation energies down to a few tens of keV \cite{Nesterenko2020,Hukkanen2023} can be extracted, allowing to obtain new insight into the nuclear structure. 

Masses of neutron-rich ruthenium isotopes up to $A = 116$~\cite{Hager2007,Hakala2011} have been measured before with the JYFLTRAP double Penning trap mass spectrometer~\cite{Eronen2012}. However, for the cases where long-lived isomers are present, namely $^{113,115}$Ru, the time-of-flight ion-cyclotron-resonance (TOF-ICR) \cite{Konig1995} technique used at that time did not provide enough resolving power to separate the ground- and isomeric-states in $^{113}$Ru or to detect the isomer in $^{115}$Ru unknown at that time. Therefore these results might have suffered from a systematic shift for the reported ground-state mass-excess values~\cite{Hukkanen2023}. More exotic ruthenium isotopes were studied using the Experimental Storage Ring at GSI~\cite{Knoebel2016}. However, $^{117}$Ru had the uncertainty 2.4-times increased by the Atomic Mass Evaluation 2020 (AME20) evaluators while the mass-excess value of $^{118}$Ru was rejected due to a significant 700-keV deviation from the mass trends~\cite{Huang2021}.

In this work, we report on the direct mass measurement of the ground states of  $^{113,115,117}$Ru and the isomeric states in $^{113}$Ru and $^{115}$Ru, the latter being the shortest-lived state ($T_{1/2} = 76(6)$~ms) ever measured at JYFLTRAP so far. The role of deformation for the systematics of masses in this region and the nature of the isomeric state in $^{115}$Ru are analysed within the context of the recent global microscopic models BSkG1~\cite{Scamps2021} and BSkG2~\cite{Ryssens2022,Ryssens2023} that are based on self-consistent Hartree-Fock-Bogoliubov (HFB) calculations using a Skyrme energy density functional (EDF).

\section{\label{sec:exp}Experimental method}

The masses of neutron-rich ruthenium isotopes were studied at the Ion Guide Isotope Separator On-Line (IGISOL) facility~\cite{Moore2013} using the JYFLTRAP double Penning trap mass spectrometer~\cite{Eronen2012} during two experiments. The isotopes of interest were produced in proton-induced fission by impinging a 25-MeV proton beam onto a thin target, $^{232}$Th for $^{113}$Ru and $^{\rm nat}$U for $^{115,117}$Ru. First, the fission fragments were stopped in a helium gas cell operating close to 300~mbar from which they were extracted and guided using a sextupole ion guide~\cite{Karvonen2008}. Then, the produced ions were accelerated to $30q$~keV and mass-separated based on their mass-to-charge ratio using a 55 degree dipole magnet. The continuous mass-separated beam was cooled and bunched using the helium buffer gas-filled radio-frequency quadrupole cooler-buncher~\cite{Nieminen2001}. Finally, the ion bunches were injected into the JYFLTRAP double Penning trap.

In the first trap of JYFLTRAP, known as the purification trap, the ion bunch was cooled, centered and the ions of interest were selected utilizing the mass-selective buffer gas cooling technique \cite{Savard1991}. After that, the purified ion sample was sent into the second trap, called the precision trap, where the mass measurements took place. 

In addition, $^{113}$Ru$^{2+}$ ions were produced via the in-trap decay of $^{113}$Tc ($T_{1/2} = 152(8)$~ms \cite{NUBASE20}). The $^{113}$Tc$^+$ ions, produced via fission, were captured in the first trap after which, the ion motion was let to cool for $102$~ms. Then a dipolar excitation on the magnetron frequency was applied for $10$~ms. During the trapping time a fraction of the $^{113}$Tc$^+$ ion sample $\beta$-decay to $^{113}$Ru$^{2+}$. Quadrupolar excitation of $100$~ms was used to select the ions of interest by matching the excitation frequency of $^{113}$Ru$^{2+}$ ions. After, the  $^{113}$Ru$^{2+}$ ions were sent to the second trap for the precision mass measurement. 

In the presence of a magnetic field of strength $B$, the mass $m$ of an ion is related to its cyclotron frequency $\nu_c$:

\begin{equation}
\nu_c = \frac{1}{2\pi}\frac{q}{m}B \mathrm{,}
\end{equation}
where $q/m$ is the charge-to-mass ratio of the measured ion. To determine the magnetic field strength precisely, $^{133}$Cs$^+$ ions from the IGISOL offline surface ion source station \cite{Vilen2020} were used as a reference for the mass measurement of $^{113,115}$Ru$^+$ ground states and $^{117}$Ru$^+$. For the mass measurement of isomeric states in $^{113,115}$Ru, the ground-state masses were used as a reference. To account for the temporal magnetic field fluctuations, ruthenium ions and their references were measured alternately. The atomic mass $m$ is determined from the frequency ratio $r=\nu_{\rm c,ref.}/\nu_{c}$ between the reference ions and the ions of interest:
\begin{equation} \label{eq:mass}
M = \frac{z_i}{z_{ref}} (M_{ref} - z_{ref} m_{e}) r + z_i m_{e},
\end{equation}
where $M_{ref}$ is an atomic mass of the reference, $m_{e}$ is an electron mass, $z_i$ and $z_{ref}$ are charge states of the ion of interest and the reference ion, respectively. The isomer excitation energies were extracted as follows:
\begin{equation}
\label{eq:Ex}
E_{x} = (r-1)[m_{\rm gs} - z m_e]c^2 \mathrm{,}
\end{equation} 
where $m_{\rm gs}$ is the ground-state atomic mass, $z$ is the charge state of the reference ion and the ion of interest (both either singly-charged, $z=1$, or doubly-charged, $z=2$ in this experiment), and $c$ is the speed of light in vacuum. Contribution from electron binding energies are on the order of eV and have thus been neglected.

\begin{figure}
\includegraphics[width=\columnwidth]{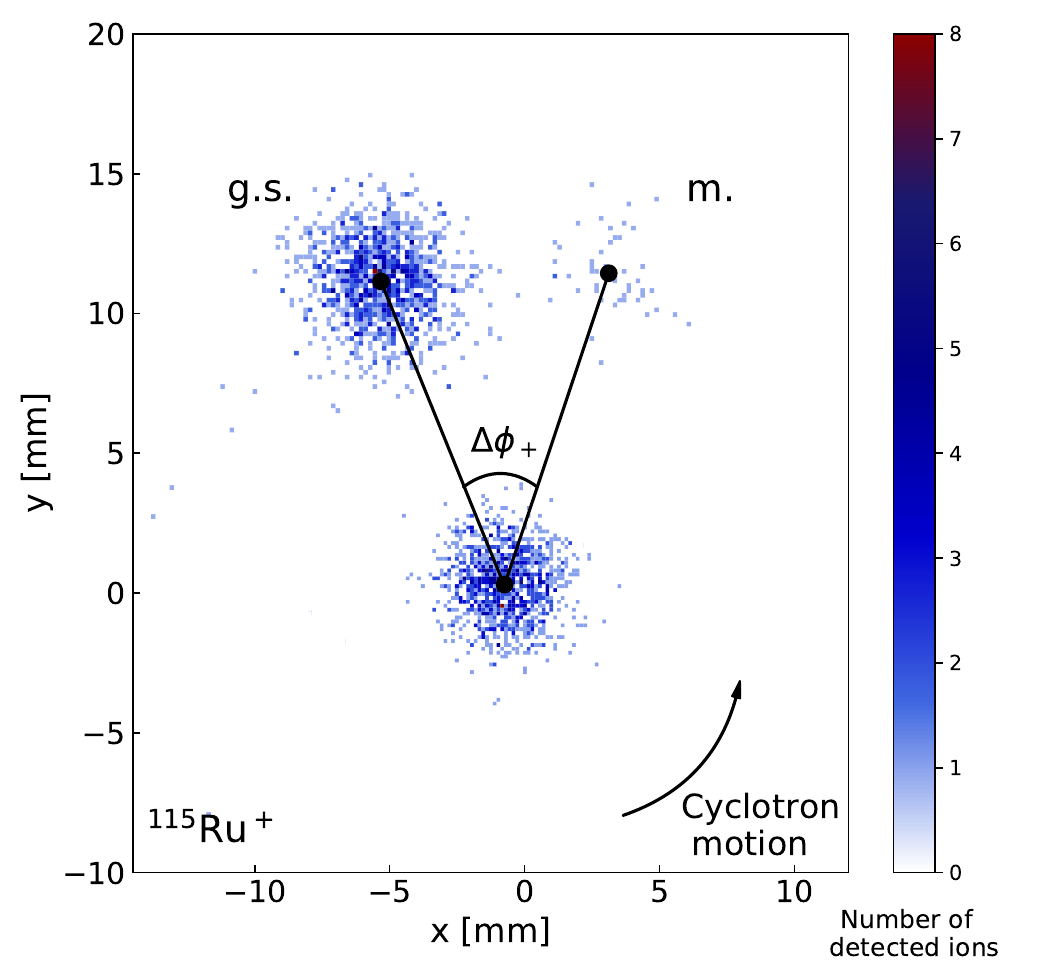}
\caption{\label{fig:115ru_pi-icr} A PI-ICR measurement of $^{115}$Ru ground state versus the isomeric state with a 100 ms accumulation time. Only the projection of cyclotron motion on the 2D MCP is shown. The angle difference $\Delta\phi_+$ leads to an excitation energy of 129(5) keV. The center spot, i.e. without any excitation, is also shown.}
\end{figure}

To measure the masses of the ground- and isomeric states in $^{113,115}$Ru, the PI-ICR technique~\cite{Eliseev2014,Nesterenko2018} was utilized in the precision trap. With the PI-ICR technique, the ion's cyclotron frequency is determined by measuring the sum of the accumulated residual phases of the magnetron ($\phi_-$) and cyclotron ($\phi_+$) motions that are projected onto a position-sensitive microchannel plate (2D MCP) detector after a phase accumulation time $t_{acc}$. Using the polar angles of the cyclotron ($\alpha_+$) and magnetron ($\alpha_-$) phase images on the detector, the angle between the motion phases with respect to the center spot is $\alpha_c=\alpha_+-\alpha_-$. This can be used for the determination of the cyclotron frequency:
\begin{equation}
\nu_c = \frac{\alpha_c + 2\pi n}{2 \pi t_{acc}},
\end{equation}
with $n$ being the sum of full revolutions performed at the magnetron and modified cyclotron frequencies during the phase accumulation time in the precision trap. 
We used the following accumulation times for the PI-ICR mass measurements:
$557$~ms for the $^{113}$Ru$^+$ ground and isomeric state, $220$~ms for the $q = 2+$ ions of $^{113}$Ru isomeric state, $200$~ms for the $^{115}$Ru$^+$ ground state and 100 ms for the $^{115}$Ru$^+$ isomer (see Fig.~\ref{fig:115ru_pi-icr}). The measurement pattern utilized at JYLFTRAP is described in more detail in Refs.~\cite{Nesterenko2018,Nesterenko2021} and the PI-ICR measurement technique in Ref. \cite{Eliseev2014}. 

\begin{figure}
\includegraphics[width=\columnwidth]{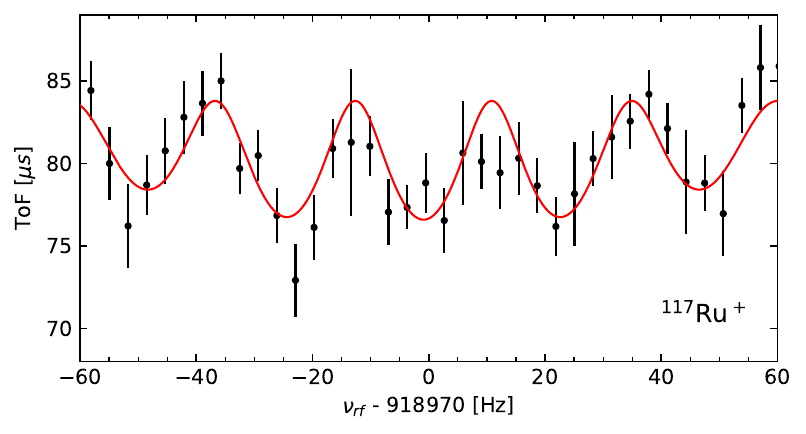}
\caption{\label{fig:117ru_tof-icr} A TOF-ICR measurement of $^{117}$Ru$^+$ using a 10-30-10~ms (on-off-on) Ramsey excitation pattern. The mean data points are shown in black, the fit of the theoretical curve \cite{Kretzschmar2007} in red.}
\end{figure}

For $^{117}$Ru$^+$, the TOF-ICR technique~\cite{Graff1980, Konig1995} was applied. The ion's cyclotron frequency $\nu_c$ in TOF-ICR technique is determined from a time-of-flight resonance measured with the 2D MCP detector, located outside the strong magnetic field of the trap. To enhance the resolving power, the Ramsey method of time-separated oscillatory fields \cite{Kretzschmar2007, George2007} was utilised. A short 10-30-10~ms (on-off-on) pattern was used in order to minimize the decay losses (see Fig.~\ref{fig:117ru_tof-icr}). 

In the mass measurement of $^{113}$Ru and $^{115}$Ru, the ground state and the isomer were in the precision trap at the same time. It is known that when two or more ions of different masses are present in the trap simultaneously, the ion-ion interaction can cause a frequency shift~\cite{Nesterenko2021}. To account for the ion-ion interaction, a count-rate class analysis~\cite{Kellerbauer2003,Nesterenko2021}, i.e. analysis of the variation of the frequency ratio with the number of ions stored simultaneously in the trap, was performed for the ground state ion of $^{115}$Ru, while for other cases it was not statistically feasible. 
At JYFLTRAP the systematic uncertainty related to temporal magnetic field fluctuation has been determined to be $\delta B/B = 2.01(25) \times 10^{-12}$ min$^{-1}$ $\times \delta t$ \cite{Nesterenko2021}, where $\delta t$ is the time between the measurements. In all of the measurements the maximum systematic uncertainty related to the temporal magnetic field fluctuations was calculated but was found to be negligible compared to the statistical uncertainty. 
We added a further mass-dependent uncertainty of $\delta_m r/r = -2.35(81) \times 10^{-10} / \textnormal{u} \times (m_{\rm ref} - m)$ and a residual systematic uncertainty of $\delta_{\rm res}r/r=9\times 10^{-9}$ for measurements where the $A/q$ for the reference and ion-of-interest were not the same, i.e. when using the $^{133}$Cs ions as reference~\cite{Nesterenko2021}. A systematic uncertainty related to the magnetron phase advancement and systematic angle error were also accounted for in the PI-ICR measurements. A more detailed description on the systematic uncertainties and their determination at JYFLTRAP can be found in Ref.~\cite{Nesterenko2021}.

\section{\label{sec:results}Results}

The ground- and isomeric-state mass of $^{113,115}$Ru and the ground-state mass of $^{117}$Ru are reported in detail below. The measured frequency ratios ($r$), mass-excess values (ME) and excitation energies ($E_x$) are summarized in Table~\ref{tab:results}. 

\begin{table*}
\caption{\label{tab:results} The measured frequency ratios ($r=\nu_{c, \rm ref.}/\nu_{c}$) and corresponding mass-excess values determined in this work (ME) using the listed reference ions (Ref.). The charge state $z$ used both for the reference ions and the ion of interest is also listed. The reported uncertainties are total uncertainties. The mass-excess values from the AME20~\cite{AME20} and NUBASE20~\cite{NUBASE20} (ME$_{\rm lit.}$) and the differences ${\mathrm{Diff.} = \mathrm{ME}-\mathrm{ME}_{\rm lit.}}$ are given for comparison. All the half-lives $T_{1/2}$ and spin-parity assignments $J^{\pi}$ of $^{113,113m,117}$Ru are taken from the NUBASE2020 evaluation~\cite{NUBASE20} while spin-parity assignments for $^{115}$Ru$^{\rm gs,m}$ are taken from Ref.~\cite{Rissanen2011} and this work. \#~denotes that the spin is based on systematics while parentheses indicate a tentative assignment.}
\begin{ruledtabular}
\begin{tabular}{lllllllllll}
 Nuclide & $T_{1/2}$ (ms) & $J^{\pi}$ & Ref. & $z$ & $r$ & ME (keV) & ME$_{lit.}$ (keV) & $E_x$ (keV)  & $E_{x, lit.}$ (keV) & Diff. (keV)\\ \hline
$^{113}$Ru & 800(50) & $(1/2^{+})$ & $^{133}$Cs & 1 & \num{0.849647289(12)} & \num{-71874.6(15)} & \num{-71870(40)} & & & \num{-5(40)}\\
$^{113}$Ru$^{m}$ & 510(30) & $(7/2^{-})$ & $^{113}$Ru & 1 & \num{1.000000951(10)}\footnotemark[1] & \num{-71774.6(18)}  &  & \num{100.0(11)}  &  & \\
 &  &  &  $^{113}$Ru & 2 & \num{1.000000963(14)}\footnotemark[2]  & \num{-71773.3(21)} &  & \num{101.3(15)} &  & \\
\multicolumn{6}{r}{Final value:}  & \num{-71774.2(17)}  & \num{-71740(50)}  & \num{100.5(8)}\footnotemark[3] & \num{131(33)} & \num{-34(50)}\\
 \noalign{\vskip 1.3mm}  
 $^{115}$Ru & 318(19) & $(3/2)^{+}$ & $^{133}$Cs & 1  & \num{0.864742653(23)} & \num{-66054.7(28)} & \num{-66105(25)} & & & \num{50(26)}\\
 $^{115}$Ru$^{m}$ & 76(6) & $(9/2)^{-}$ & $^{115}$Ru & 1 & \num{1.000001206(47)} & \num{-65925.6(58)} & \num{-66110(90)} & \num{129(5)} & \num{82(6)} & \num{184(91)}\\
  \noalign{\vskip 1.3mm}  
 $^{117}$Ru & 151(3) & $3/2^+$\# & $^{133}$Cs & 1 & \num{0.87984374(52)} & \num{-59526(64)} & \num{-59490(430)} & & & \num{-36(435)} \\
\end{tabular}
\footnotetext[1]{Measured with $1^+$ ions produced directly in fission.}
\footnotetext[2]{Measured with $2^+$ ions produced in in-trap-decay of $^{113}$Tc.}
\footnotetext[3]{Weighted average of the two measurements.}
\end{ruledtabular}
\end{table*}

\subsection{$^{113}$Ru}
\label{sec:113Ru}

The ground-state mass excess of $^{113}$Ru, $-71874.6(15)$~keV, was determined using $^{133}$Cs$^+$ ions as a reference. The isomer excitation energy, $E_x =100.5(8)$~keV, was determined against the ground state, both as singly-charged ions produced directly in fission as well as doubly-charged ions produced via in-trap decay of $^{113}$Tc$^+$ (for details see Sect.~\ref{sec:exp}). This yields a mass excess of $-71774.2(17)$~keV for the isomer. 

\begin{figure}[h!t!b]
\includegraphics[clip, trim=0.0cm 0cm 1.5cm 1.0cm,width=\columnwidth]{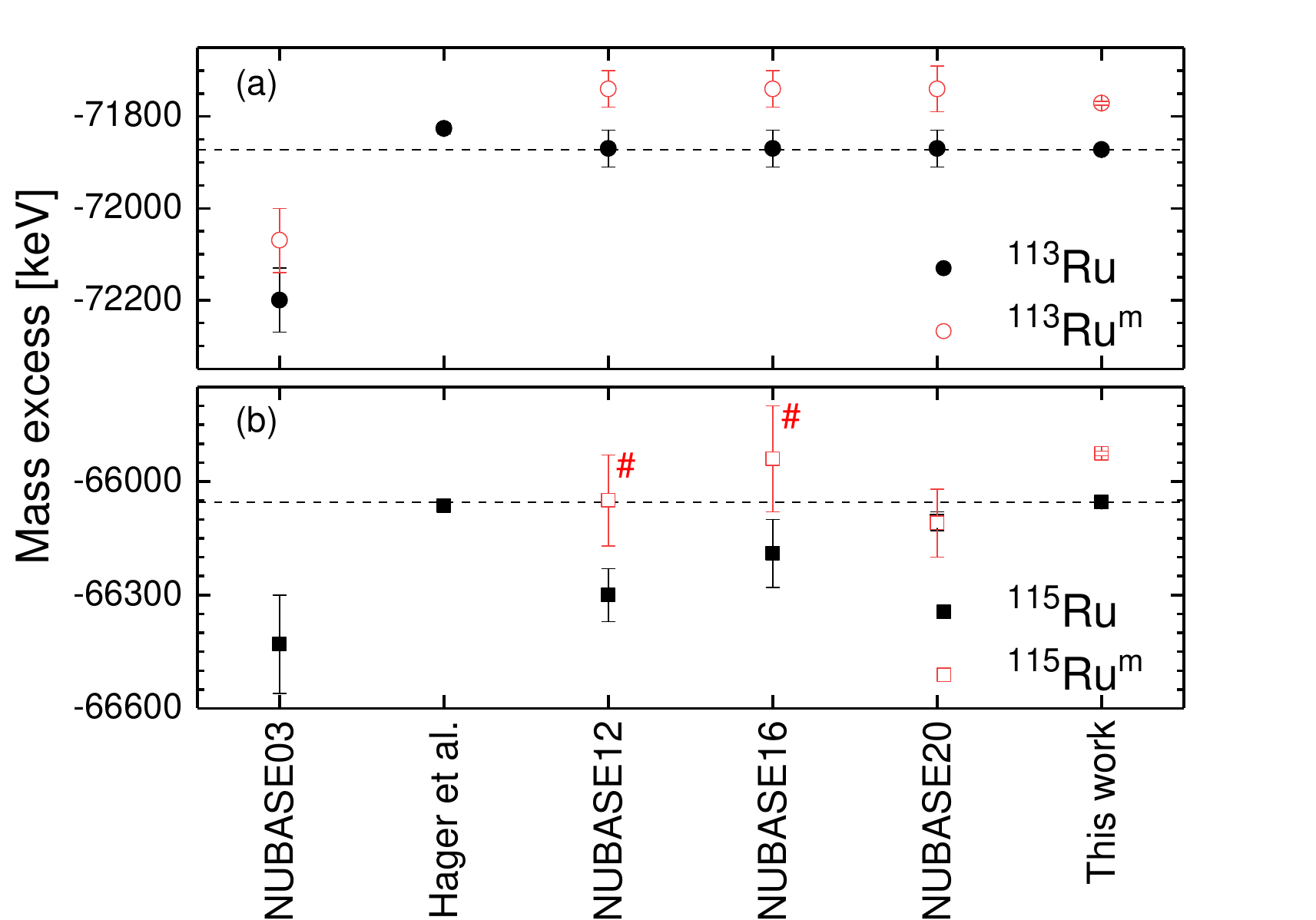}
\caption{\label{fig:115ru_nubase} The mass-excess values determined in this work for the ground states (solid black symbols) and isomers (open red symbols) in (a) $^{113}$Ru and (b) $^{115}$Ru, in comparison with the revised JYFLTRAP value reported by Hager \textit{et al.} \cite{Hager2007,AME20} and different NUBASE compilations \cite{NUBASE03,NUBASE12,NUBASE16,NUBASE20}. The dashed black lines show the ground-state mass-excess values determined in this work. \#~denotes mass-excess values based on systematics.}
\end{figure}

The mass of $^{113}$Ru has been previously measured at JYFLTRAP by Hager et al. \cite{Hager2007}, using the TOF-ICR technique with a 400-ms quadrupolar excitation time and $^{105}$Ru$^+$ ions as a reference. With the AME20~\cite{AME20} mass value for $^{105}$Ru, this results in a mass-excess value of $-71826(12)$~keV. The revised value is in between the ground- and isomeric-state mass-excess values reported in this work (see Fig.~\ref{fig:115ru_nubase}.(a)) suggesting that a mixture of states was measured in Ref. \cite{Hager2007}. A similar effect was observed in Rh isotopes, as reported in Ref.~\cite{Hukkanen2023}.

\begin{figure}[h!t!b]
\includegraphics[width=0.7\columnwidth]{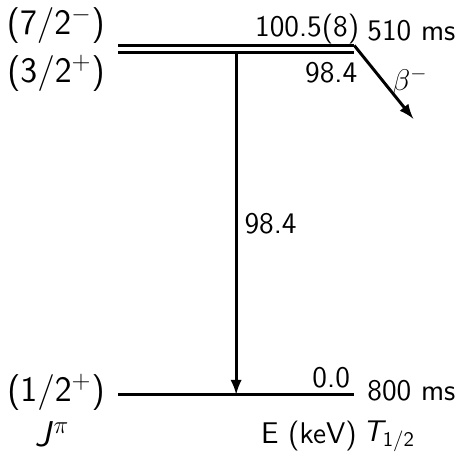}
\caption{\label{fig:113ru_excitedstates}Proposed partial level scheme of $^{113}$Ru based on this work and Ref.~\cite{Blachot2010}.}
\end{figure}

The reported mass-excess values are in agreement with the NUBASE20 evaluation~\cite{NUBASE20} where it was correctly assumed that the value measured in Ref.~\cite{Hager2007} was a mixture of the ground state and an isomer at 131(33) keV. To date, the isomeric-state excitation energy was not based on direct experimental observations but on the suggestion that it has to lie in between the states at 98 and 164 keV in $^{113}$Ru \cite{Kurpeta1998,Kurpeta2007}. In this work, we have confirmed this hypothesis by determining the excitation energy for the first time and by placing the isomer just above the 98-keV state (see Fig. \ref{fig:113ru_excitedstates}). The production of both long-lived states in $^{113}$Ru in the $\beta$-decay of $^{113}$Tc is also in agreement with the work by Kurpeta et al. \cite{Kurpeta1998}.

\subsection{$^{115}$Ru}
\label{sec:115Ru}

The ground state mass excess, $-66054.7(28)$~keV, was measured against a $^{133}$Cs$^+$ reference. The isomer excitation energy, $E_x = 129(5)$~keV, was determined against the ground state resulting in a mass excess of $-65925.6(58)$~keV for the isomer. 

Our ground-state mass excess value is in agreement with the previous TOF-ICR-based JYFLTRAP measurement (ME = $-66064.0(69)$ \cite{Hager2007,AME20}) after adjusting for the updated mass of the reference $^{120}$Sn ion. In our previous work we have observed that for nuclei with low-lying isomeric states the masses obtained with the TOF-ICR method are a weighted average of the ground state and the isomer masses \cite{Hukkanen2023}. In the case of $^{115}$Ru, an apparent absence of the isomer influence on the measured mass can be explained by a relatively short half-life of the isomeric state ($T_{1/2} = 76$~ms \cite{NUBASE20}) compared to the 300~ms excitation time used in Ref. \cite{Hager2007}. 

Figure~\ref{fig:115ru_nubase}.(b) shows a comparison of our measurement with the values reported in NUBASE evaluations on nuclear and decay properties from 2003 \cite{NUBASE03}, 2012 \cite{NUBASE12}, 2016 \cite{NUBASE16} and 2020 \cite{NUBASE20} as well as the revised JYFLTRAP value of Ref.~\cite{Hager2007}. Changes between different editions of NUBASE can be explained as due to varying input data. In NUBASE03 \cite{NUBASE03}, the only entry for $^{115}$Ru was from a $\beta$-decay end-point energy study \cite{Kratz2000}. 
After the JYFLTRAP measurement by Hager et al. \cite{Hager2007}, a long-lived isomeric state in $^{115}$Ru was discovered \cite{Kurpeta2010}, and the evaluators of NUBASE12 \cite{NUBASE12} applied a special procedure for mixtures of isomeric states assuming the excitation energy to be 250(100) keV. In NUBASE16 \cite{NUBASE16}, the $\beta$-decay end-point energy study was excluded from the global fit and the only remaining information was from Ref. \cite{Hager2007}. Finally, in NUBASE20 \cite{NUBASE20}, the energy of the isomeric state was adjusted to 82(6)~keV based on the value originally proposed in Ref. \cite{Kurpeta2010}. However, the isomeric-state excitation energy seems not to be taken into account for the mass-excess value of the isomer but only for its uncertainty.

\subsection{$^{117}$Ru}

The value determined in this work, $-59526(64)$~keV, is in agreement with AME20~\cite{AME20} and it is almost seven times more precise. The mass-excess value adopted in AME20, $-59490(430)$~keV \cite{AME20}, is based on storage-ring measurements \cite{Matos2004,Knoebel2016} but with the uncertainty artificially increased by evaluators \cite{Huang2021}. The only known isomeric state has a half-life of $2.49(6)~\mu$s \cite{NUBASE20} which is much shorter than the measurement cycle used in this work.

\section{\label{sec:disc} Discussion}

In this section, we discuss the experimental results and compare them to the BSkG-family of
models of nuclear structure~\cite{Scamps2021,Ryssens2022,Ryssens2023}. This section
is organised as follows: we first establish the theoretical framework in 
Sec.~\ref{sec:disc:theory} and then proceed to study first the trends of the
ground state (g.s.) binding energies of neutron-rich Ru isotopes in Sec.~\ref{sec:disc:gs}.
Sec.~\ref{sec:disc:isomer} discusses the isomeric state in $^{115}$Ru as well
as the implication of our measurement of its excitation energy.

\subsection{Theoretical framework}
\label{sec:disc:theory}

The BSkG-family of models responds to the need for reliable data on 
   the structural properties of exotic nuclei in different fields of research
   and in astrophysics in particular. 
   These models are based on an empirical Energy Density Functional (EDF)
   of Skyrme type that models the effective in-medium nucleon-nucleon
   interaction. The concept of an EDF allows for a global yet microscopic
   description of all relevant quantities at a reasonable computational cost. 
   The coupling constants of the EDF are the main element of 
   phenomenology in this type of model and have to be adjusted to experimental 
   data. Since binding energies are crucial ingredients for
   the modeling of nuclear reactions, the ensemble of known nuclear masses is
   a key ingredient of the parameter adjustment of the BSkG models. Because of 
   this, these models reach root-mean-square (rms) deviations better than 800
   keV on the thousands of masses included in AME20~\cite{AME20}. This performance
   is not at all competitive with the uncertainties of the 
   measurements we report on here, but it nevertheless reflects the state-of-the-art
   in global mass modeling: it is only matched by some of the older BSk models
   that were adjusted in the same spirit~\cite{goriely2016}, microscopic-macroscopic
   approaches~\cite{Moller2016} and empirical models~\cite{duflo_1995}. 
   The latter two types of model become particularly accurate when refined 
   with machine learning techniques~\cite{niu_2022}, but either do not extend 
   their predictions to other observables or struggle to describe them with the 
   same parameter values deduced from the masses.

The BSkG-family so far counts two entries: BSkG1~\cite{Scamps2021} and 
BSkG2~\cite{Ryssens2022,Ryssens2023}. Both models combine a description of 
many hundreds of measured charge radii and realistic predictions for the 
properties of infinite nuclear matter with a description of the AME20 masses 
with similar accuracy (rms deviations of 741 and 678 keV, respectively).
Although some of the BSk models reach an rms deviation below 600 keV~\cite{goriely2016}, 
BSkG1 and BSkG2 are better adapted to study the neutron-rich Ru isotopes as they
rely on a three-dimensional representation of the nucleus, thereby accomodating naturally the triaxial deformation that 
is known to be particularly relevant for this region of the nuclear chart. 
BSkG2 incorporates a full treatment of the so-called `time-odd' 
terms in an EDF~\cite{Ryssens2022} and improves systematically on the description of 
fission properties compared to its predecessor~\cite{Ryssens2023}. Since (i) 
the inclusion of the time-odd terms did not result in a meaningful improvement of 
our global description of binding energies and (ii) fission properties are not directly
related to the masses, \textit{a priori} we expect BSkG1 and BSkG2 to be of 
roughly equal quality for the task at hand and therefore we will compare experiment to both models in what follows.

Large-scale EDF-based models of nuclear structure such as the BSk- and BSkG-models 
describe the nucleus in terms of one single product wavefunction, typically 
of the Bogoliubov type. The simplicity of such an \textit{ansatz}, as compared to the 
complexity of the many-body problem, is compensated for by allowing for spontaneous
symmetry breaking in the mean fields. By considering such deformed configurations EDF-based models  
can account for a large part of the effects of nuclear collectivity on bulk
properties such as masses while remaining at the mean-field level and thus
keeping calculations tractable. Nevertheless, symmetry breaking comes at 
considerable computational cost. 
For all calculations that we report on, we 
employed the MOCCa code~\cite{Ryssens2016} to represent the single-nucleon 
wavefunctions on a three-dimensional coordinate mesh. All numerical parameters such as the 
mesh point spacing are identical to those employed in the adjustment of both
BSkG models~\cite{Scamps2021,Ryssens2022}. 

In a three-dimensional calculation, the quadrupole deformation of a nucleus of 
mass $A$ can be described by way of the (dimensionless) deformation $\beta_{2}$ 
and the triaxiality angle $\gamma$, defined as
\begin{align}
\beta_2 &= \frac{4 \pi}{ 3 R^2 A }\sqrt{Q_{20}^2 + 2 Q_{22}^2}\, , \\ 
\gamma &= \text{atan} \left( \sqrt{2} Q_{22}/Q_{20} \right)\, ,
\end{align}
where $R = 1.2 A^{1/3}$ fm. The quadrupole moments $Q_{20}$ and $Q_{22}$ are
defined in terms of integrals of the total nuclear density and spherical 
harmonics, see for instance Ref.~\cite{Scamps2021}. Axially symmetric prolate 
and oblate shapes correspond to $\gamma = 0^{\circ}$ and $60^{\circ}$, 
respectively, while intermediate values of the triaxiality angle in between those 
two extremes indicate triaxial shapes.  

\begin{figure}
\centering
\includegraphics[width=\columnwidth]{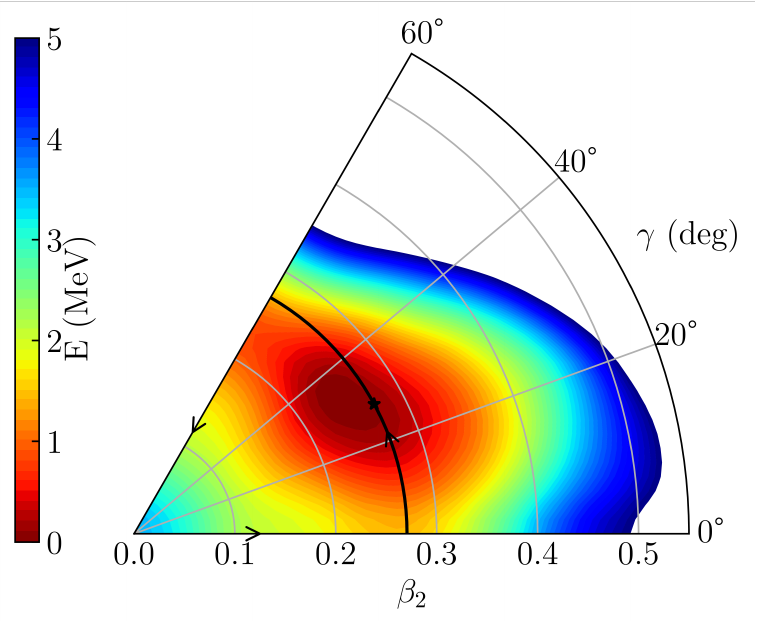}
\caption{\label{fig:pes}Potential energy surface in the $(\beta,\gamma)$ - plane
         for false-vacuum calculations (see text) of $^{115}$Ru with BSkG2. 
         The trajectory followed by the Nilsson diagram in Fig.~\ref{fig:nilsson} is 
         indicated by black arrows. The location of the minimum obtained in a 
         complete calculation of $^{115}$Ru is indicated by a black star.}
\end{figure}

We show in Fig.~\ref{fig:pes} 
the potential energy surface (PES) of $^{115}$Ru  in the $\beta-\gamma$ plane as
obtained with BSkG2, calculations with BSkG1 leading to a similar PES. Since 
$^{115}$Ru has an odd number of nucleons, 
Fig.~\ref{fig:pes} shows the result of so-called `false-vacuum' calculations, 
where we constrained the expected number of neutrons to $\langle N \rangle = 71$, 
but otherwise treated the nucleus as if it were even-even. We emphasize that 
all the calculations for which we report masses do not rely on this approximation: 
for both BSkG1 and BSkG2 our treatment of the odd-mass Ru isotopes includes
self-consistent blocking of a neutron quasiparticle. 
For BSkG2, we also include the energy contribution of the finite spin and 
current densities induced by the presence of the odd neutrons. 
For more details on our treatment of odd-mass and odd-odd nuclei, see the discussion 
in Ref.~\cite{Ryssens2022}. A complete calculation for $^{115}$Ru that includes
blocking leads to the deformation shown as a black star on Fig.~\ref{fig:pes}; 
its offset with respect to the minimum of the false-vacuum calculations is
due to the polarisation induced by the odd neutron.

Qualitatively, the false-vacuum PES of $^{115}$Ru looks similar to the PES
of $^{112}$Rh that we discussed in Ref.~\cite{Hukkanen2023}: we observe a somewhat 
broad triaxial minimum near $\gamma = 30^{\circ}$ of significant quadrupole
deformation. Close inspection reveals some quantitative differences:
$\beta_2 \sim 0.27$ is here somewhat smaller than the value $0.3$ obtained
for $^{112}$Rh for instance. Another difference is the energy gain due to triaxiality: 
the difference between the oblate saddle point and the minimum on Fig.~\ref{fig:pes}
is about 800 keV, while it exceeds 1 MeV for $^{112}$Rh.
This can be linked to the four additional neutrons in $^{115}$Ru compared to $^{112}$Rh: as 
we approach the $N=82$ shell closure, the neutrons have less freedom to exploit
quadrupole correlations and the importance of (static) quadrupole 
deformation in general and triaxial deformation in particular diminishes.

\subsection{The g.s. masses of Ru isotopes and their trends}
\label{sec:disc:gs}

For the chain of Ru isotopes between $N=65$ and $N=73$, 
BSkG1 reproduces the absolute g.s. binding energies best: the deviation with respect to experiment for the 
absolute mass excesses averages to 360 keV and never exceeds 640 keV. The performance
of BSkG2 is not as good: an average deviation of 650 keV with a deviation 
of up to $1.175$ MeV for $^{115}$Ru. Interestingly, the sign of the deviation 
is consistent: both models overbind these Ru isotopes and hence produce
mass excesses that are too large in absolute size. As discussed before, the experimental uncertainties  
are several orders of magnitude beyond the accuracy of global models like
BSkG1 and BSkG2: instead of comparing 
absolute masses in more detail, we will focus in what follows primarily on the trends of 
mass differences.

\begin{figure}
\includegraphics[width=\columnwidth]{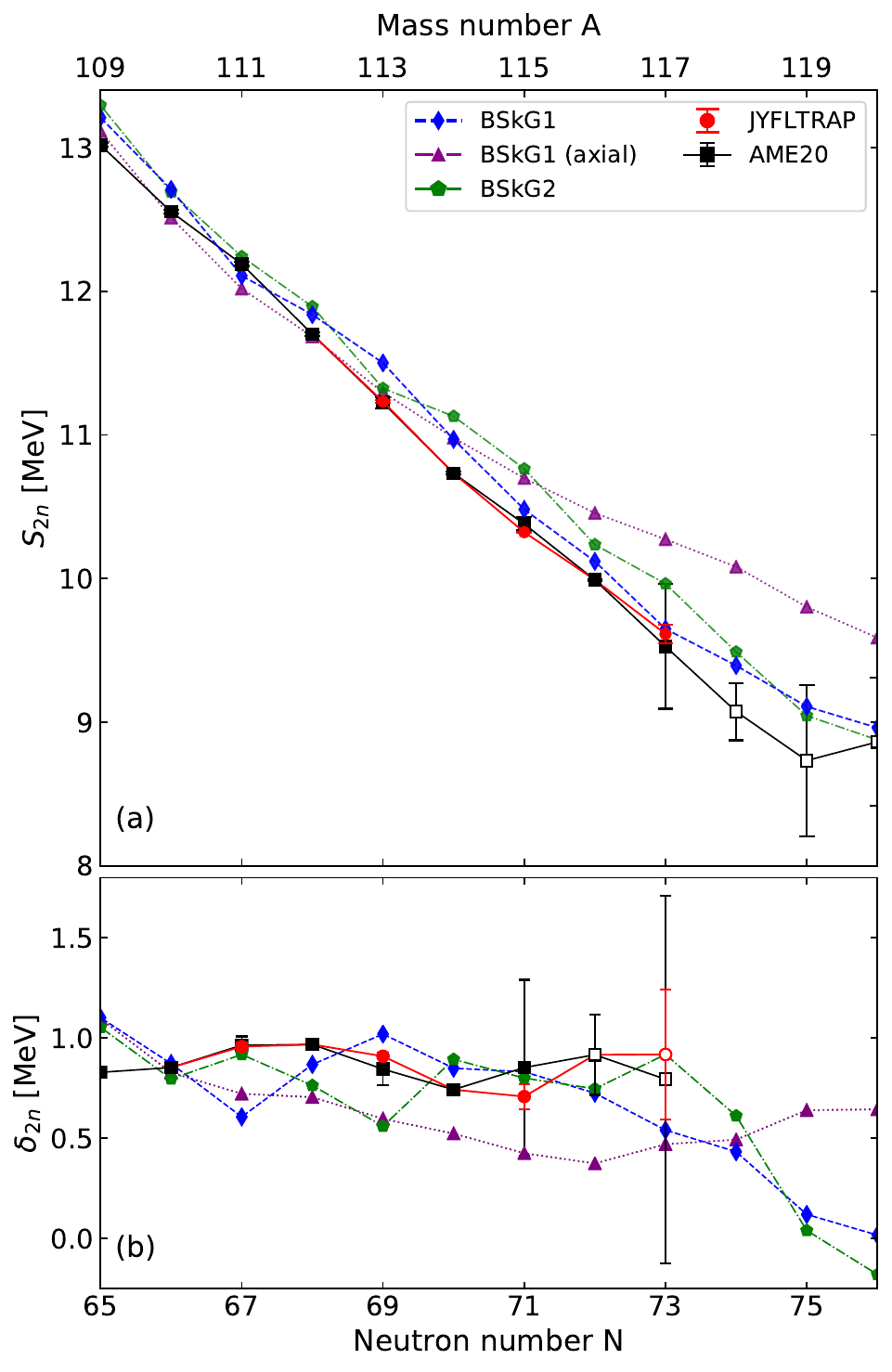}\\
\caption{\label{fig:S2N_JYFLTRAP+AME2020} 
Comparison of mass differences along the Ru isotopic chain: experimental
values either as tabulated in AME20 (black squares) or updated with the 
results of our new measurements (red circles) versus calculated values obtained with BSkG1 (blue diamonds), BSkG1 axial (purple triangles, see text) and BSkG2 (green pentagons). Open markers for the experimental results represent values at least partially 
based on extrapolated mass values from AME20 \cite{AME20}.
Top panel: two-neutron separation energies $S_{2n}$. Bottom panel: two-neutron
shell gaps $\delta_{2n}$.
}
\end{figure}

We start with the two-neutron separation energy $S_{2n}$, defined as:
\begin{equation}
S_{2n}(Z,N) = ME(Z,N-2) - ME(Z,N) +2ME(0,1) \mathrm{,}
\end{equation}
where $ME(Z,N)$ is the mass excess of a nucleus with $Z$ protons and $N$ neutrons and $ME(0,1)$ is the mass excess of the neutron. 
The top panel of Fig.~\ref{fig:S2N_JYFLTRAP+AME2020} compares the $S_{2n}$ values
derived from the newly measured masses to the values reported in the 
AME20~\cite{AME20} evaluation and the two mass models. We also show the results 
of the less general calculations with BSkG1 reported on in Ref.~\cite{Scamps2021,Hukkanen2023}, 
which restrict the nucleus to axially symmetric configurations.

For the less exotic $^{109,111,113}$Ru, all three calculations with BSkG-models
reproduce the general trend of the experimental $S_{2n}$ rather well, although
deviations on the order of several hundred keV are clearly visible. 
For the BSkG1 model, the description of the more neutron-rich isotopes follows 
the trend of the more stable ones, systematically overestimating the $S_{2n}$ values by a small value. 
BSkG2 also overestimates the separation energies and describes their overall
trend, but with deviations that are somewhat larger than those of its predecessor.  
Calculations with BSkG1 that are restricted to axial shapes, however, 
entirely miss the experimental trend.

We can furthermore discuss the slope of the $S_{2n}$ curve by introducing the 
empirical two-neutron shell gaps $\delta_{2n}$:
\begin{equation}
\delta_{2n}(Z,N) = S_{2n}(Z,N) - S_{2n}(Z,N+2) \mathrm{,} 
\end{equation}
which we show in the bottom panel of Fig.~\ref{fig:S2N_JYFLTRAP+AME2020}. The 
new JYFLTRAP measurement for $^{115}$Ru clearly establishes that the 
slope of the $S_{2n}$ in this isotopic chain evolves smoothly at least 
until $N=71$. Although the corresponding curves are less regular, the BSkG1 and 
BSkG2 results produce $\delta_{2n}$ values that remain close to experiment
up to $N=71$. For the heavier $N=72$, $73$ and $74$ isotopes, whose
experimental $\delta_{2n}$ values are at least partially based on extrapolated
AME20 values, the two models predict no major change in slope either. It 
is only for $N=75-76$ that 
BSkG1 and BSkG2 predict a change in slope that is correlated with the 
disappearance of triaxial deformation for $N\geq 76$. For $^{120}$Ru and even 
more neutron-rich isotopes, the models predict axially symmetric prolate shapes 
with deformation that gradually diminishes towards $N=82$.

Finally, we discuss the three-point neutron gaps $\Delta^{(3)}_n(Z,N)$:
\begin{equation}
\begin{split}
\Delta_{n}^{(3)}(Z,N) =& \frac{(-1)^{N}}{2} \big[ME(Z,N+1)\\&+ME(Z,N-1)-2ME(Z,N)\big] \mathrm{.}
\end{split}
\end{equation}
This quantity estimates the average distance between the curves that interpolate
the masses of the even-$N$ and odd-$N$ isotopes, respectively, as a function of 
neutron number. It is particularly sensitive to the neutron pairing, but it can also be
affected by variations in the structure of these isotopes with $N$. The new 
experimental results confirm the continuation of the trend of less
   exotic isotopes: the three-point gaps for the even-$N$ isotopes at $N=66$, $68$, $70$, and $72$
   are all equal within error-bars. For $N=70$, our new result actually brings
   the $\Delta^{(3)}_n$ value more in line with this trend. The updated value of 
   $\Delta_n^{(3)}$ for $N=71$ falls significantly out of the uncertainty
   range of AME20, which reflects the lack of accuracy of the AME20 estimate
   for the excitation energy of the isomeric state of $^{115}$Ru. Nevertheless, it is not
   dramatically larger than the gap values for $N=69$ and $N=71$.

The BSkG2  model generally overestimates $\Delta^{(3)}_n$ and its curve exhibits
features at $N=68$, $69$, and $70$ that are not seen in the experimental data. 
   BSkG1 on the other hand, provides a fair description of the experimental results, whether 
    including or not triaxial deformation. Yet even this model is clearly not 
    without flaws: the deviation of the full calculation w.r.t.~experiment grows 
   with $N$ from $N=69$ onwards. In this respect, the deviation
   between the calculated BSkG1 value and the updated point at $N=73$ (which
   incorporates the recommended AME20 binding energy for $^{118}$Ru) seems 
   ominous. We note in passing that both BSkG models systematically 
   overestimate $\Delta^{(3)}_n$ along odd-$Z$ isotopic chains, which we
   discovered for the first time during the study of neighbouring Rh isotopes in 
   Ref.~\cite{Hukkanen2023}. Similarly, both models overestimate
   the calculated three-point proton gaps in odd-$N$ isotopic chains. The common
   origin of these issues is the failure of both models to account for a small 
   amount of binding energy in odd-odd nuclei that is usually ascribed to the
   residual interaction between the two odd nucleons, see Ref.~\cite{Ryssens2022}.  
   This issue does not affect our discussion here, but it explains why both 
   models describe much better the three-point neutron gaps in even-$Z$ Ru 
   isotopes than in odd-$Z$ Rh isotopes.

\begin{figure}
\includegraphics[width=\columnwidth]{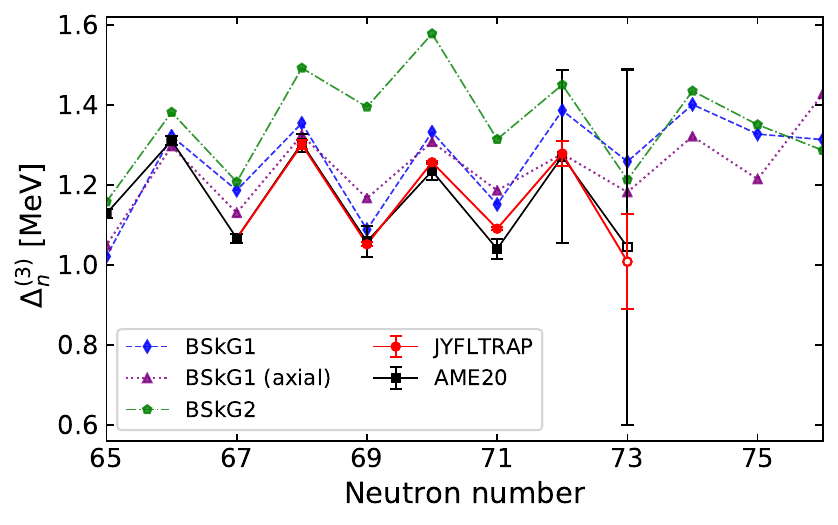}
\caption{\label{fig:oes} Same as for Fig.~\ref{fig:S2N_JYFLTRAP+AME2020}, but for the three-point neutron gap $\Delta_n^{\rm (3)}$.}
\end{figure}

We have established that performance of BSkG2 for the $N=65-71$ Ru isotopes 
   is worse than that of BSkG1 for absolute masses as well as all mass differences
   discussed. Since these models are the result of a complicated parameter adjustment which
   is global in scope, it is hard to pinpoint a particular source of this
   (local) deficiency.  As we remarked in the previous section, we did not
   a priori expect that BSkG2 would offer an improved description of the 
   measured masses. Although the difference we observe between models
   indicates BSkG1 as the tool of choice for future studies of this region, this
   does not imply that BSkG2 is a step backwards compared to its predecessor. 
   The newer model presents a different compromise on the very large number of
   observables included in the parameter adjustment, leading to a worse 
   description of the nuclei we study here but also to an improved description
   of other observables~\cite{Ryssens2022}.

To close this section, we note again that our new measurement indicates a 
rather uneventful continuation to $N=71$ of the trends of binding energies and 
mass differences as established for less exotic isotopes. This can be interpreted
as experimental confirmation that the structural evolution of nuclei in this
isotopic chain is smooth rather than dramatic. From the point of view of the 
BSkG models this was expected: from $N=55$ onwards, the Ru isotopes exhibit
triaxial deformation that smoothly evolves with neutron number until $N=76$. 
The authors of Ref.~\cite{Rissanen2011} relied on the Woods-Saxon 
single-particle spectrum of Ref.~\cite{Xu_2002} to interpret the change in 
(tentative) ground state spin assignment in $^{113-115}$Ru (($1/2^+$) and ($3/2^+$), 
respectively) as a sign of a shape transition from prolate to oblate 
deformation. The trend of masses and mass differences does not seem to support
such scenario.

\subsection{\label{sec:disc:isomer} The isomer in $^{115}$Ru}

The isomeric state in $^{115}$Ru was reported for the first time in Ref.~\cite{Kurpeta2010}, discussing the analysis of a $\beta$ decay experiment. The authors observed that the 61.7-keV $\gamma$ ray is not in coincidence with a $\beta$ particle or any other $\gamma$ ray. In addition, the half-life extracted from this transition, $T_{1/2} = 76(6)$~ms, differed from the half-life obtained for the $^{115}$Ru ground-state ($T_{1/2} = 318(19)$~ms). Consequently, it was assumed that the isomeric state de-excites via an unobserved $\gamma$ ray having energy below Ru K x-rays ($E \approx 20$~keV), which we label $\gamma_1$, followed by an emission of the 61.7-keV $\gamma$ ray, labeled as $\gamma_2$. 

With the assumption of the energy of $\gamma_1$ being below 20~keV, the observed ruthenium K x-rays were associated solely with the emission of K internal conversion electrons from the $\gamma_2$ transition. This observation enabled a determination of the $\gamma_2$ K-internal conversion coefficient ($\alpha_K = 2.7(6)$ \cite{Kurpeta2010}) by calculating the ratio of the ruthenium K x-rays and the $\gamma_2$ transitions. 

\begin{figure}[h!t!b]
\includegraphics[width=0.7\columnwidth]{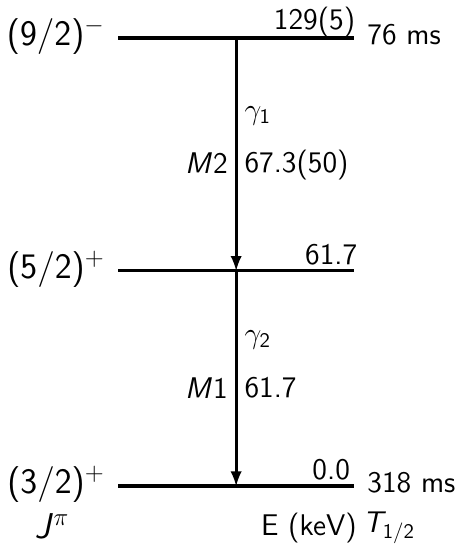}
\caption{\label{fig:115ru_excitedstates} Proposed level scheme of $^{115}$Ru based on this work and Refs.~\cite{Kurpeta2010,Rissanen2011}.}
\end{figure}

\begin{figure*}
\includegraphics[width=\textwidth]{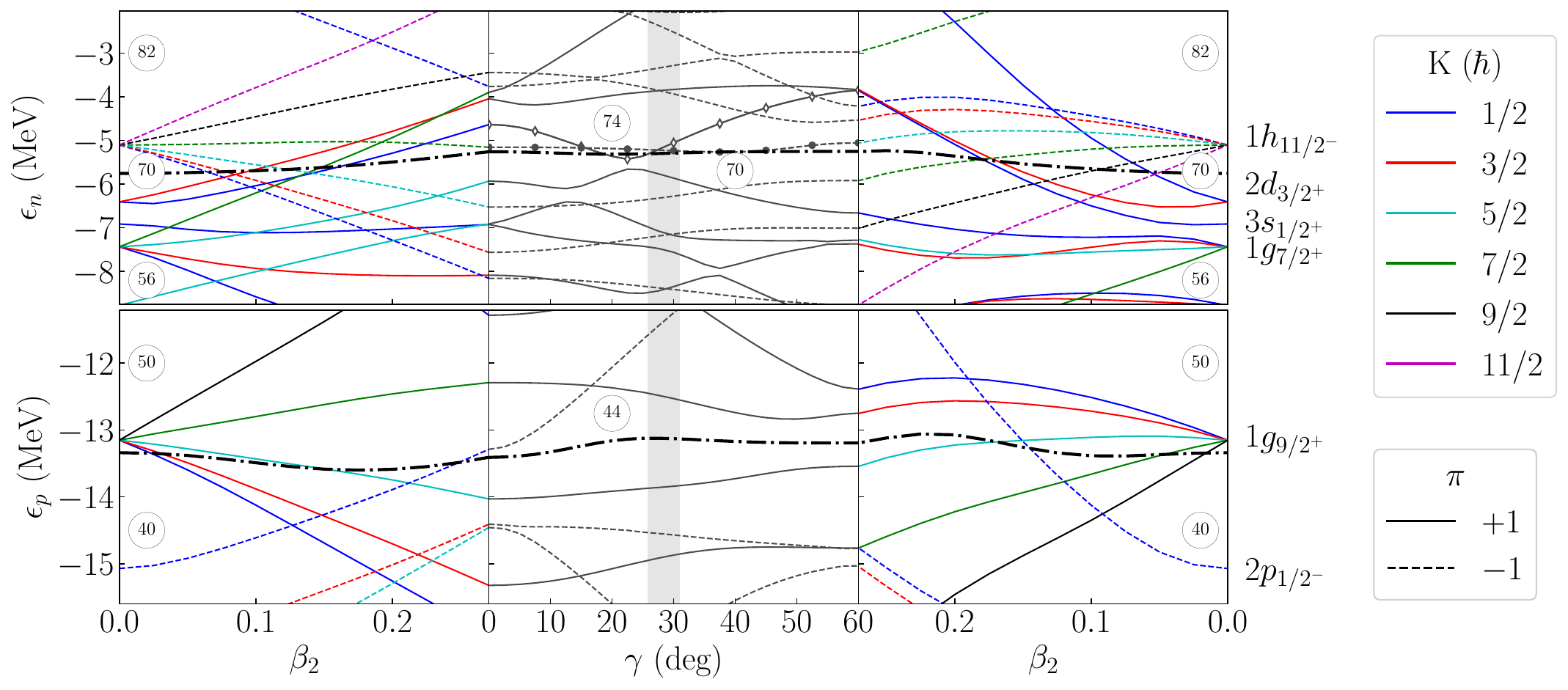}
\caption{\label{fig:nilsson} (Color online) Eigenvalues of the single-particle 
Hamiltonian for neutrons (top row) and protons (bottom row) along the path in the $\beta-\gamma$ plane 
indicated by arrows in Fig.~\ref{fig:pes} for $^{115}$Ru (see text for details). 
The Fermi energy is drawn as a dash-dotted line, while full (dashed) lines 
indicate single-particle levels of positive (negative) parity. The three 
indicated regions correspond to axially symmetric prolate shape with (left column), fixed total quadrupole deformation $\beta_2=0.27$ with varying $\gamma$ (center column) and axially oblate shape with $\gamma=60^\circ$ (right column). The vertical gray band in the center panels is centered at $\gamma = 28.4^{\circ}$, the value obtained in a complete, i.e. blocked, calculation of $^{115}$Ru. The quantum numbers of the shells at sphericity are indicated on the right-hand-side. Two-neutron levels near the Fermi energy are highlighted by markers in the middle column: these are the positive and negative parity levels referred to in the text, respectively, as $|\diamond\rangle$ and $|\bullet\rangle$.}
\end{figure*}

The new isomer excitation energy reported in this work renders previous calculations incorrect. However, if one assumes that (i) the total intensity ($\gamma$ rays and internal conversion electrons emission) of $\gamma_1$ and $\gamma_2$ is identical, (ii) $\gamma_1$ has a pure $M2$ character and (iii) $\gamma_2$ has a pure $M1$ character, the observed ratio of the ruthenium K x-rays to $\gamma_2$ would be equal to 2.8(8). Any other assumptions regarding the multipolarity of both transitions would lead to a ratio that differs significantly from the experimental value of 2.7(6)~\cite{Kurpeta2010}. Therefore, we propose $M2$ and $M1$ multipolarities for $\gamma_1$ and $\gamma_2$, respectively. By assigning $(3/2)^+$ as the ground-state spin-parity as proposed in~\cite{Rissanen2011} from a detailed $\beta$-decay spectroscopy experiment of $^{115}$Ru, a tentative $(9/2)^-$ isomer assignment can be adopted, see Fig.~\ref{fig:115ru_excitedstates}. 


A precise description of the level scheme of $^{115}$Ru is beyond the 
capabilities of current large-scale models such as BSkG1 and BSkG2, but we
can use them to gain a qualitative understanding of the existence of the isomeric
state. To this end, we show in Fig.~\ref{fig:nilsson}, the Fermi energy and the 
single-particle energies for both neutrons and protons obtained in false-vacuum 
calculations for $^{115}$Ru with BSkG2 along the trajectory in the $\beta-\gamma$ 
plane indicated by the arrows in Fig.~\ref{fig:pes}. Although symmetry-breaking
allows models such as BSkG1 and BSkG2 to grasp a significant part of the effect 
of collectivity on nuclear structure, here is where we pay the price: we can 
no longer use the quantum numbers of an operator associated with a broken symmetry to label
single-particle states. At the spherical point, on the utmost left and right of 
Fig.~\ref{fig:nilsson}, no symmetry is broken and all single-particle levels 
are simultaneous eigenstates of three operators with three associated quantum
numbers: the angular momentum squared $\hat{J}^2$ with quantum number $J$, 
parity $\hat{P}$ with quantum number $\pi$ and the $z$-component of the angular 
momentum $\hat{J}_z$ with quantum number $K$. The quantum numbers of the orbitals
at the spherical point are indicated in the traditional spectroscopic notation 
on the right of Fig.~\ref{fig:nilsson}. Along the first segment of the path on
Fig.~\ref{fig:pes}, we break rotational symmetry but conserve axial symmetry: 
the levels in the left-most column are no longer eigenstates of $\hat{J}^2$ 
but retain the $K$ quantum number\footnote{For axially symmetric configurations, 
we always align the symmetry axis with the z-axis in the simulation volume.}, 
which is indicated by colors on Fig.~\ref{fig:pes}. When exploring finite values 
of $\gamma$ along the second segment of the path on Fig.~\ref{fig:pes} axial 
symmetry is broken and $K$ can no longer be used to label the single-particle 
states, hence the absence of colors in the middle column of Fig.~\ref{fig:nilsson}. 
The final segment of the path explores oblate shapes which are axially 
symmetric, such that levels in the right column of Fig.~\ref{fig:nilsson} can 
again be color-coded. For all our calculations we conserve parity, such that
$\pi$ is a good single-particle quantum number along the entire path that we 
can use to distinguish between levels of positive (full lines) and negative parity 
(dashed lines) in all columns of Fig.~\ref{fig:nilsson}.

This loss of single-particle quantum numbers also translates to the many-body 
state: the BSkG-models cannot currently offer definite angular momentum assignments 
for calculated ground states for odd-mass and odd-odd nuclei. Doing so would 
require symmetry-restoration techniques \cite{Bally21a} whose 
application is presently still out of the scope of global models for reasons of
their numerical cost and because of formal issues with the type of EDF assumed 
for the BSkG models. We are however not entirely without options: we can 
calculate expectation values $\langle i| \hat{J_z} | i \rangle$, which will not 
be half-integer multiples of $\hbar$ but which nevertheless tell us something 
about the angular momentum of the single-particle state $|i\rangle$. In the 
limit of a non-interacting particle-core model of the ground states of
odd-mass Ru isotopes, the angular momentum expectation value of the odd
neutron will also be the expectation value of angular momentum of the many-body
state.

We discussed a qualitatively similar Nilsson diagram obtained for $^{112}$Rh in 
Ref.~\cite{Hukkanen2023} and repeat here a few observations that are common
to both nuclei before discussing the isomer. Local minima in the PES correspond 
to deformations for which the single-particle level density near the Fermi energy
is low: for nuclei with $Z=43$, $44$, and $45$, the protons drive the appearance of triaxial deformation since their single-particle spectrum at 
$\beta_2 \sim 0.28-0.3, \gamma \sim 30^{\circ}$ is very sparse. In this region 
of the PES only positive parity states orbital are near the Fermi energy,
matching the parity assignments of all even-$N$ Tc and Rh isotopes. The 
single-particle level density of the neutrons on the other hand is much 
higher, resulting in a closely-spaced set of levels with different parities
near the Fermi energy.
  
We interpret the close interleaving of positive and negative parity neutron states
with different angular momentum content as the origin of the isomeric state
in $^{115}$Ru. Two neutron states are nearly degenerate near the Fermi 
energy at the location of the minimum of the PES: these are highlighted in 
the middle column of Fig.~\ref{fig:nilsson} and we will refer to them by their
markers: $|\diamond\rangle$ and $|\bullet\rangle$ .
These levels differ in their parity, 
but also in their angular momentum content: near $\gamma = 30^{\circ}$
the positive parity state has an average $\langle \diamond | \hat{J}_z | \diamond \rangle \approx 0.73 \hbar$, 
while that of the negative parity state is significantly larger, 
$\langle \bullet | \hat{J}_z | \bullet \rangle \approx 4.13 \hbar$.
Since the odd-neutron can be assigned to each of these levels, we expect the 
appearance of two low-lying levels with opposite parity in the spectrum of 
$^{115}$Ru that are close in energy yet differ substantially in their angular 
momentum, hence one of them being an isomer.
    Finally, the $(5/2)^+$ state in between the g.s. and the isomer on 
    Fig.~\ref{fig:115ru_excitedstates} could be rotational in character: 
    taking the calculated moments of inertia of $^{115}$Ru and under the 
    assumption of a rigid triaxial rotor, a $1\hbar$ change in total angular
    momentum corresponds to about $88$ keV of excitation energy.

Moving beyond simple arguments based on a non-interacting particle-core 
picture and the Nilsson diagram, we 
explicitly calculated the lowest-lying configuration of each parity in $^{115}$Ru
with both BSkG1 and BSkG2. One of these is the calculated g.s.~, whose binding energy figured in 
the previous section: for BSkG1 this is the state with positive parity and for 
BSkG2 this is the one with negative parity. In both cases, we find an excited 
state of opposite parity at low excitation energy; 33 keV and 90 keV for BSkG1 
and BSkG2 respectively. For BSkG2, we have direct access to the average many-body 
angular momentum along the z-axis: a small value $\langle J_z \rangle \approx 0.7 \hbar$ 
for the positive parity state and a large one, $\langle J_z \rangle \approx 3.1 \hbar$, 
for the negative parity state. These calculations support our conclusions drawn from
    the Nilsson diagram and the calculated excitation energy are very roughly 
    comparable to the experimental isomer excitation energy. These results should
    not be overinterpreted: all relevant energy differences are very small and
    the neutron spectrum in Fig.~\ref{fig:nilsson} is very complicated. Small 
    changes to any aspect of the model will affect the precise location of level
    crossings and therefore the ordering of levels. Our calculated excitation energies should thus not be taken 
    as a precise prediction, but rather as a confirmation that two states of 
    opposite parity that differ little in energy can be constructed with different
    angular momentum content. Predicting their ordering and energy 
    difference with accuracy is beyond BSkG1 and BSkG2, or for that matter, any
    large-scale model that we are aware of.

The same mechanism can be used to interpret the isomerism in nearby $N=71$ isotones: isomeric states with half-lives on the order of seconds or longer have been observed in $^{116}$Rh, $^{118}$Ag and $^{119}$Cd whereas shorter-lived isomeric states are known in $^{114}$Tc and $^{117}$Pd~\cite{NUBASE20}.
    For $Z=42-46$, one can expect from Fig.~\ref{fig:nilsson} triaxial 
    deformation with a sparse proton single-particle spectrum and two
    low-lying states arising from neutron orbitals of different parities.
    The experimental systematics extend much further: 
    in the entire range of $Z=43-57$, low-lying isomers have been 
    observed~\cite{NUBASE20}. A more in-depth study of isomerism in the $N=71$ 
    isotones would certainly require more diagrams like Fig.~\ref{fig:nilsson}
    for larger proton numbers and is outside of the scope of this study. 
    Nevertheless, we remark that 
    both BSkG1 and BSkG2 predict triaxial deformation for almost all $N=71$ 
    isotones in the range $Z=40-60$\footnote{The only exceptions occur
    for BSkG1 near the $Z=50$ shell closure: $^{118}$Ag, $^{119}$Cd, $^{120}$In 
    and $^{121}$Sn remain axially symmetric.}.
    
%

\section{\label{sec:summary}Summary}

The masses of $^{113,115,117}$Ru have been measured using the Penning-trap mass spectrometry at the JYFLTRAP double Penning trap. The ground- and isomeric states in $^{113,115}$Ru have been separated and masses measured using the PI-ICR technique. The isomer excitation energies were determined directly for the first time. The high-precision measurement reported in this work place the $(7/2)^-$ isomeric state in $^{113}$Ru at 100.5(8)~keV, just above the $(3/2^+)$ level at 98.4(3)~keV \cite{Blachot2010}, but still in agreement with the previous prediction of 133(33)~keV \cite{NUBASE20}. For $^{115m}$Ru, the excitation energy was found to be 129(5)~keV, which is significantly larger than proposed in Ref.~\cite{Kurpeta2010} or the value listed in the most recent NUBASE evaluation, 82(6)~keV \cite{NUBASE20}.

The determined ground-state masses of $^{113,117}$Ru are in excellent agreement with the atomic mass evaluation \cite{AME20}. For $^{115}$Ru, we report a mass-excess value which is 50(26)~keV larger than reported in AME20 \cite{AME20}. However, it is in agreement with the previous JYFLTRAP mass measurement by Hager et al. \cite{Hager2007}. With the mass values determined in this work, the trend in the two-neutron separation energies continues smoothly. 

The experimental results have been compared with the global BSkG1~\cite{Scamps2021} and BSkG2~\cite{Ryssens2022,Ryssens2023} models, which allow for triaxially deformed shapes. Detailed calculations were performed for the structure of $^{115}$Ru. In the predicted triaxial deformation, the proton single-particle spectrum was found to be sparse and the predicted low-lying states arise from neutron orbitals with different parities. More systematic studies on the isomeric states in this triaxially deformed region would be needed to shed more light on the reasons for the isomerism in these nuclei. 

\begin{acknowledgments}
The present research benefited from computational resources made available on the Tier-1 supercomputer of the F\'ed\'eration Wallonie-Bruxelles, infrastructure funded by the Walloon Region under the grant agreement No 1117545. W.R. is a Research Associate of the F.R.S.-FNRS (Belgium). Work by M.B.\ has been supported by the Agence Nationale de la Recherche, France, Grant No.~19-CE31-0015-01 (NEWFUN). Funding from the European Union’s Horizon 2020 research and innovation programme under grant agreements No. 771036 (ERC CoG MAIDEN) and No. 861198–LISA–H2020-MSCA-ITN-2019 are gratefully acknowledged. M.H. acknowledges financial support from the Ellen \& Artturi Nyyss\"onen foundation. We are grateful for the mobility support from Projet International de Coop\'eration Scientifique Manipulation of Ions in Traps and Ion sourCes for Atomic and Nuclear Spectroscopy (MITICANS) of CNRS. We acknowledge the support from the Academy of Finland projects No. 295207, 306980, 327629, and 354968. J.R. acknowledges financial support from the Vilho, Yrj\"o and Kalle V\"ais\"al\"a Foundation.
\end{acknowledgments}

\bibliographystyle{apsrev}
\bibliography{biblio}

\begin{thebibliography}{49}
\expandafter\ifx\csname natexlab\endcsname\relax\def\natexlab#1{#1}\fi
\expandafter\ifx\csname bibnamefont\endcsname\relax
  \def\bibnamefont#1{#1}\fi
\expandafter\ifx\csname bibfnamefont\endcsname\relax
  \def\bibfnamefont#1{#1}\fi
\expandafter\ifx\csname citenamefont\endcsname\relax
  \def\citenamefont#1{#1}\fi
\expandafter\ifx\csname url\endcsname\relax
  \def\url#1{\texttt{#1}}\fi
\expandafter\ifx\csname urlprefix\endcsname\relax\def\urlprefix{URL }\fi
\providecommand{\bibinfo}[2]{#2}
\providecommand{\eprint}[2][]{\url{#2}}

\bibitem[{\citenamefont{Garrett et~al.}(2022)\citenamefont{Garrett, Zielińska,
  and Clément}}]{Garrett2022}
\bibinfo{author}{\bibfnamefont{P.~E.} \bibnamefont{Garrett}},
  \bibinfo{author}{\bibfnamefont{M.}~\bibnamefont{Zielińska}},
  \bibnamefont{and} \bibinfo{author}{\bibfnamefont{E.}~\bibnamefont{Clément}},
  \bibinfo{journal}{Progress in Particle and Nuclear Physics}
  \textbf{\bibinfo{volume}{124}}, \bibinfo{pages}{103931}
  (\bibinfo{year}{2022}), ISSN \bibinfo{issn}{0146-6410},
  \urlprefix\url{https://www.sciencedirect.com/science/article/pii/S0146641021000922}.

\bibitem[{\citenamefont{Srebrny et~al.}(2006)\citenamefont{Srebrny, Czosnyka,
  Droste, Rohozi\'nski, Pr\'ochniak, Zaj\c{a}c, Pomorski, Cline, Wu, B\"acklin
  et~al.}}]{Srebrny2006}
\bibinfo{author}{\bibfnamefont{J.}~\bibnamefont{Srebrny}},
  \bibinfo{author}{\bibfnamefont{T.}~\bibnamefont{Czosnyka}},
  \bibinfo{author}{\bibfnamefont{C.}~\bibnamefont{Droste}},
  \bibinfo{author}{\bibfnamefont{S.}~\bibnamefont{Rohozi\'nski}},
  \bibinfo{author}{\bibfnamefont{L.}~\bibnamefont{Pr\'ochniak}},
  \bibinfo{author}{\bibfnamefont{K.}~\bibnamefont{Zaj\c{a}c}},
  \bibinfo{author}{\bibfnamefont{K.}~\bibnamefont{Pomorski}},
  \bibinfo{author}{\bibfnamefont{D.}~\bibnamefont{Cline}},
  \bibinfo{author}{\bibfnamefont{C.}~\bibnamefont{Wu}},
  \bibinfo{author}{\bibfnamefont{A.}~\bibnamefont{B\"acklin}},
  \bibnamefont{et~al.}, \bibinfo{journal}{Nuclear Physics A}
  \textbf{\bibinfo{volume}{766}}, \bibinfo{pages}{25} (\bibinfo{year}{2006}),
  ISSN \bibinfo{issn}{03759474},
  \urlprefix\url{https://linkinghub.elsevier.com/retrieve/pii/S0375947405012029}.

\bibitem[{\citenamefont{Möller et~al.}(2006)\citenamefont{Möller, Bengtsson,
  Gillis~Carlsson, Olivius, and Ichikawa}}]{Moller2006}
\bibinfo{author}{\bibfnamefont{P.}~\bibnamefont{Möller}},
  \bibinfo{author}{\bibfnamefont{R.}~\bibnamefont{Bengtsson}},
  \bibinfo{author}{\bibfnamefont{B.}~\bibnamefont{Gillis~Carlsson}},
  \bibinfo{author}{\bibfnamefont{P.}~\bibnamefont{Olivius}}, \bibnamefont{and}
  \bibinfo{author}{\bibfnamefont{T.}~\bibnamefont{Ichikawa}},
  \bibinfo{journal}{Phys. Rev. Lett.} \textbf{\bibinfo{volume}{97}},
  \bibinfo{pages}{162502} (\bibinfo{year}{2006}),
  \urlprefix\url{https://doi.org/10.1103/PhysRevLett.97.162502}.

\bibitem[{\citenamefont{Scamps et~al.}(2021)\citenamefont{Scamps, Goriely,
  Olsen, Bender, and Ryssens}}]{Scamps2021}
\bibinfo{author}{\bibfnamefont{G.}~\bibnamefont{Scamps}},
  \bibinfo{author}{\bibfnamefont{S.}~\bibnamefont{Goriely}},
  \bibinfo{author}{\bibfnamefont{E.}~\bibnamefont{Olsen}},
  \bibinfo{author}{\bibfnamefont{M.}~\bibnamefont{Bender}}, \bibnamefont{and}
  \bibinfo{author}{\bibfnamefont{W.}~\bibnamefont{Ryssens}},
  \bibinfo{journal}{The European Physical Journal A}
  \textbf{\bibinfo{volume}{57}}, \bibinfo{pages}{333} (\bibinfo{year}{2021}),
  ISSN \bibinfo{issn}{1434-601X},
  \urlprefix\url{https://doi.org/10.1140/epja/s10050-021-00642-1}.

\bibitem[{\citenamefont{Hager et~al.}(2006)\citenamefont{Hager, Eronen, Hakala,
  Jokinen, Kolhinen, Kopecky, Moore, Nieminen, Oinonen, Rinta-Antila
  et~al.}}]{Hager2006}
\bibinfo{author}{\bibfnamefont{U.}~\bibnamefont{Hager}},
  \bibinfo{author}{\bibfnamefont{T.}~\bibnamefont{Eronen}},
  \bibinfo{author}{\bibfnamefont{J.}~\bibnamefont{Hakala}},
  \bibinfo{author}{\bibfnamefont{A.}~\bibnamefont{Jokinen}},
  \bibinfo{author}{\bibfnamefont{V.}~\bibnamefont{Kolhinen}},
  \bibinfo{author}{\bibfnamefont{S.}~\bibnamefont{Kopecky}},
  \bibinfo{author}{\bibfnamefont{I.}~\bibnamefont{Moore}},
  \bibinfo{author}{\bibfnamefont{A.}~\bibnamefont{Nieminen}},
  \bibinfo{author}{\bibfnamefont{M.}~\bibnamefont{Oinonen}},
  \bibinfo{author}{\bibfnamefont{S.}~\bibnamefont{Rinta-Antila}},
  \bibnamefont{et~al.}, \bibinfo{journal}{Phys. Rev. Lett.}
  \textbf{\bibinfo{volume}{96}}, \bibinfo{pages}{042504}
  (\bibinfo{year}{2006}),
  \urlprefix\url{https://doi.org/10.1103/PhysRevLett.96.042504}.

\bibitem[{\citenamefont{Naimi et~al.}(2010)\citenamefont{Naimi, Audi, Beck,
  Blaum, B\"ohm, Borgmann, Breitenfeldt, George, Herfurth, Herlert
  et~al.}}]{Naimi2010}
\bibinfo{author}{\bibfnamefont{S.}~\bibnamefont{Naimi}},
  \bibinfo{author}{\bibfnamefont{G.}~\bibnamefont{Audi}},
  \bibinfo{author}{\bibfnamefont{D.}~\bibnamefont{Beck}},
  \bibinfo{author}{\bibfnamefont{K.}~\bibnamefont{Blaum}},
  \bibinfo{author}{\bibfnamefont{C.}~\bibnamefont{B\"ohm}},
  \bibinfo{author}{\bibfnamefont{C.}~\bibnamefont{Borgmann}},
  \bibinfo{author}{\bibfnamefont{M.}~\bibnamefont{Breitenfeldt}},
  \bibinfo{author}{\bibfnamefont{S.}~\bibnamefont{George}},
  \bibinfo{author}{\bibfnamefont{F.}~\bibnamefont{Herfurth}},
  \bibinfo{author}{\bibfnamefont{A.}~\bibnamefont{Herlert}},
  \bibnamefont{et~al.}, \bibinfo{journal}{Phys. Rev. Lett.}
  \textbf{\bibinfo{volume}{105}}, \bibinfo{pages}{032502}
  (\bibinfo{year}{2010}),
  \urlprefix\url{https://link.aps.org/doi/10.1103/PhysRevLett.105.032502}.

\bibitem[{\citenamefont{Chaudhuri et~al.}(2013)\citenamefont{Chaudhuri,
  Andreoiu, Brunner, Chowdhury, Ettenauer, Gallant, Gwinner, Kwiatkowski,
  Lennarz, Lunney et~al.}}]{Chaudhuri2013}
\bibinfo{author}{\bibfnamefont{A.}~\bibnamefont{Chaudhuri}},
  \bibinfo{author}{\bibfnamefont{C.}~\bibnamefont{Andreoiu}},
  \bibinfo{author}{\bibfnamefont{T.}~\bibnamefont{Brunner}},
  \bibinfo{author}{\bibfnamefont{U.}~\bibnamefont{Chowdhury}},
  \bibinfo{author}{\bibfnamefont{S.}~\bibnamefont{Ettenauer}},
  \bibinfo{author}{\bibfnamefont{A.~T.} \bibnamefont{Gallant}},
  \bibinfo{author}{\bibfnamefont{G.}~\bibnamefont{Gwinner}},
  \bibinfo{author}{\bibfnamefont{A.~A.} \bibnamefont{Kwiatkowski}},
  \bibinfo{author}{\bibfnamefont{A.}~\bibnamefont{Lennarz}},
  \bibinfo{author}{\bibfnamefont{D.}~\bibnamefont{Lunney}},
  \bibnamefont{et~al.}, \bibinfo{journal}{Phys. Rev. C}
  \textbf{\bibinfo{volume}{88}}, \bibinfo{pages}{054317}
  (\bibinfo{year}{2013}),
  \urlprefix\url{https://link.aps.org/doi/10.1103/PhysRevC.88.054317}.

\bibitem[{\citenamefont{Eliseev et~al.}(2013)\citenamefont{Eliseev, Blaum,
  Block, Droese, Goncharov, Minaya~Ramirez, Nesterenko, Novikov, and
  Schweikhard}}]{Eliseev2013}
\bibinfo{author}{\bibfnamefont{S.}~\bibnamefont{Eliseev}},
  \bibinfo{author}{\bibfnamefont{K.}~\bibnamefont{Blaum}},
  \bibinfo{author}{\bibfnamefont{M.}~\bibnamefont{Block}},
  \bibinfo{author}{\bibfnamefont{C.}~\bibnamefont{Droese}},
  \bibinfo{author}{\bibfnamefont{M.}~\bibnamefont{Goncharov}},
  \bibinfo{author}{\bibfnamefont{E.}~\bibnamefont{Minaya~Ramirez}},
  \bibinfo{author}{\bibfnamefont{D.~A.} \bibnamefont{Nesterenko}},
  \bibinfo{author}{\bibfnamefont{Y.~N.} \bibnamefont{Novikov}},
  \bibnamefont{and}
  \bibinfo{author}{\bibfnamefont{L.}~\bibnamefont{Schweikhard}},
  \bibinfo{journal}{Phys. Rev. Lett.} \textbf{\bibinfo{volume}{110}},
  \bibinfo{pages}{082501} (\bibinfo{year}{2013}),
  \urlprefix\url{https://link.aps.org/doi/10.1103/PhysRevLett.110.082501}.

\bibitem[{\citenamefont{Eliseev et~al.}(2014)\citenamefont{Eliseev, Blaum,
  Block, Dörr, Droese, Eronen, Goncharov, Höcker, Ketter, Minaya~Ramirez
  et~al.}}]{Eliseev2014}
\bibinfo{author}{\bibfnamefont{S.}~\bibnamefont{Eliseev}},
  \bibinfo{author}{\bibfnamefont{K.}~\bibnamefont{Blaum}},
  \bibinfo{author}{\bibfnamefont{M.}~\bibnamefont{Block}},
  \bibinfo{author}{\bibfnamefont{A.}~\bibnamefont{Dörr}},
  \bibinfo{author}{\bibfnamefont{C.}~\bibnamefont{Droese}},
  \bibinfo{author}{\bibfnamefont{T.}~\bibnamefont{Eronen}},
  \bibinfo{author}{\bibfnamefont{M.}~\bibnamefont{Goncharov}},
  \bibinfo{author}{\bibfnamefont{M.}~\bibnamefont{Höcker}},
  \bibinfo{author}{\bibfnamefont{J.}~\bibnamefont{Ketter}},
  \bibinfo{author}{\bibfnamefont{E.}~\bibnamefont{Minaya~Ramirez}},
  \bibnamefont{et~al.}, \bibinfo{journal}{Appl. Phys. B}
  \textbf{\bibinfo{volume}{114}}, \bibinfo{pages}{107} (\bibinfo{year}{2014}),
  \urlprefix\url{https://doi.org/10.1007/s00340-013-5621-0}.

\bibitem[{\citenamefont{Nesterenko et~al.}(2020)\citenamefont{Nesterenko,
  Kankainen, Kostensalo, Nobs, Bruce, Beliuskina, Canete, Eronen, Gamba,
  Geldhof et~al.}}]{Nesterenko2020}
\bibinfo{author}{\bibfnamefont{D.}~\bibnamefont{Nesterenko}},
  \bibinfo{author}{\bibfnamefont{A.}~\bibnamefont{Kankainen}},
  \bibinfo{author}{\bibfnamefont{J.}~\bibnamefont{Kostensalo}},
  \bibinfo{author}{\bibfnamefont{C.}~\bibnamefont{Nobs}},
  \bibinfo{author}{\bibfnamefont{A.}~\bibnamefont{Bruce}},
  \bibinfo{author}{\bibfnamefont{O.}~\bibnamefont{Beliuskina}},
  \bibinfo{author}{\bibfnamefont{L.}~\bibnamefont{Canete}},
  \bibinfo{author}{\bibfnamefont{T.}~\bibnamefont{Eronen}},
  \bibinfo{author}{\bibfnamefont{E.}~\bibnamefont{Gamba}},
  \bibinfo{author}{\bibfnamefont{S.}~\bibnamefont{Geldhof}},
  \bibnamefont{et~al.}, \bibinfo{journal}{Physics Letters B}
  \textbf{\bibinfo{volume}{808}}, \bibinfo{pages}{135642}
  (\bibinfo{year}{2020}), ISSN \bibinfo{issn}{0370-2693},
  \urlprefix\url{https://www.sciencedirect.com/science/article/pii/S0370269320304457}.

\bibitem[{\citenamefont{Hukkanen et~al.}(2023)\citenamefont{Hukkanen, Ryssens,
  Ascher, Bender, Eronen, Gr\'evy, Kankainen, Stryjczyk, Al~Ayoubi, Ayet
  et~al.}}]{Hukkanen2023}
\bibinfo{author}{\bibfnamefont{M.}~\bibnamefont{Hukkanen}},
  \bibinfo{author}{\bibfnamefont{W.}~\bibnamefont{Ryssens}},
  \bibinfo{author}{\bibfnamefont{P.}~\bibnamefont{Ascher}},
  \bibinfo{author}{\bibfnamefont{M.}~\bibnamefont{Bender}},
  \bibinfo{author}{\bibfnamefont{T.}~\bibnamefont{Eronen}},
  \bibinfo{author}{\bibfnamefont{S.}~\bibnamefont{Gr\'evy}},
  \bibinfo{author}{\bibfnamefont{A.}~\bibnamefont{Kankainen}},
  \bibinfo{author}{\bibfnamefont{M.}~\bibnamefont{Stryjczyk}},
  \bibinfo{author}{\bibfnamefont{L.}~\bibnamefont{Al~Ayoubi}},
  \bibinfo{author}{\bibfnamefont{S.}~\bibnamefont{Ayet}}, \bibnamefont{et~al.},
  \bibinfo{journal}{Phys. Rev. C} \textbf{\bibinfo{volume}{107}},
  \bibinfo{pages}{014306} (\bibinfo{year}{2023}),
  \urlprefix\url{https://link.aps.org/doi/10.1103/PhysRevC.107.014306}.

\bibitem[{\citenamefont{Hager et~al.}(2007)\citenamefont{Hager, Elomaa, Eronen,
  Hakala, Jokinen, Kankainen, Rahaman, Rinta-Antila, Saastamoinen, Sonoda
  et~al.}}]{Hager2007}
\bibinfo{author}{\bibfnamefont{U.}~\bibnamefont{Hager}},
  \bibinfo{author}{\bibfnamefont{V.-V.} \bibnamefont{Elomaa}},
  \bibinfo{author}{\bibfnamefont{T.}~\bibnamefont{Eronen}},
  \bibinfo{author}{\bibfnamefont{J.}~\bibnamefont{Hakala}},
  \bibinfo{author}{\bibfnamefont{A.}~\bibnamefont{Jokinen}},
  \bibinfo{author}{\bibfnamefont{A.}~\bibnamefont{Kankainen}},
  \bibinfo{author}{\bibfnamefont{S.}~\bibnamefont{Rahaman}},
  \bibinfo{author}{\bibfnamefont{S.}~\bibnamefont{Rinta-Antila}},
  \bibinfo{author}{\bibfnamefont{A.}~\bibnamefont{Saastamoinen}},
  \bibinfo{author}{\bibfnamefont{T.}~\bibnamefont{Sonoda}},
  \bibnamefont{et~al.}, \bibinfo{journal}{Phys. Rev. C}
  \textbf{\bibinfo{volume}{75}}, \bibinfo{pages}{064302}
  (\bibinfo{year}{2007}),
  \urlprefix\url{https://link.aps.org/doi/10.1103/PhysRevC.75.064302}.

\bibitem[{\citenamefont{Hakala et~al.}(2011)\citenamefont{Hakala,
  Rodr{\'i}guez-Guzm{\'a}n, Elomaa, Eronen, Jokinen, Kolhinen, Moore,
  Penttil{\"a}, Reponen, Rissanen et~al.}}]{Hakala2011}
\bibinfo{author}{\bibfnamefont{J.}~\bibnamefont{Hakala}},
  \bibinfo{author}{\bibfnamefont{R.}~\bibnamefont{Rodr{\'i}guez-Guzm{\'a}n}},
  \bibinfo{author}{\bibfnamefont{V.~V.} \bibnamefont{Elomaa}},
  \bibinfo{author}{\bibfnamefont{T.}~\bibnamefont{Eronen}},
  \bibinfo{author}{\bibfnamefont{A.}~\bibnamefont{Jokinen}},
  \bibinfo{author}{\bibfnamefont{V.~S.} \bibnamefont{Kolhinen}},
  \bibinfo{author}{\bibfnamefont{I.~D.} \bibnamefont{Moore}},
  \bibinfo{author}{\bibfnamefont{H.}~\bibnamefont{Penttil{\"a}}},
  \bibinfo{author}{\bibfnamefont{M.}~\bibnamefont{Reponen}},
  \bibinfo{author}{\bibfnamefont{J.}~\bibnamefont{Rissanen}},
  \bibnamefont{et~al.}, \bibinfo{journal}{The European Physical Journal A}
  \textbf{\bibinfo{volume}{47}}, \bibinfo{pages}{129} (\bibinfo{year}{2011}),
  ISSN \bibinfo{issn}{1434-601X},
  \urlprefix\url{https://doi.org/10.1140/epja/i2011-11129-9}.

\bibitem[{\citenamefont{Eronen et~al.}(2012)\citenamefont{Eronen, Kolhinen,
  Elomaa, Gorelov, Hager, Hakala, Jokinen, Kankainen, Karvonen, Kopecky
  et~al.}}]{Eronen2012}
\bibinfo{author}{\bibfnamefont{T.}~\bibnamefont{Eronen}},
  \bibinfo{author}{\bibfnamefont{V.}~\bibnamefont{Kolhinen}},
  \bibinfo{author}{\bibfnamefont{V.-V.} \bibnamefont{Elomaa}},
  \bibinfo{author}{\bibfnamefont{D.}~\bibnamefont{Gorelov}},
  \bibinfo{author}{\bibfnamefont{U.}~\bibnamefont{Hager}},
  \bibinfo{author}{\bibfnamefont{J.}~\bibnamefont{Hakala}},
  \bibinfo{author}{\bibfnamefont{A.}~\bibnamefont{Jokinen}},
  \bibinfo{author}{\bibfnamefont{A.}~\bibnamefont{Kankainen}},
  \bibinfo{author}{\bibfnamefont{P.}~\bibnamefont{Karvonen}},
  \bibinfo{author}{\bibfnamefont{S.}~\bibnamefont{Kopecky}},
  \bibnamefont{et~al.}, \bibinfo{journal}{Eur. Phys. J. A}
  \textbf{\bibinfo{volume}{48}}, \bibinfo{pages}{46} (\bibinfo{year}{2012}),
  \urlprefix\url{https://doi.org/10.1140/epja/i2012-12046-1}.

\bibitem[{\citenamefont{König et~al.}(1995)\citenamefont{König, Bollen,
  Kluge, Otto, and Szerypo}}]{Konig1995}
\bibinfo{author}{\bibfnamefont{M.}~\bibnamefont{König}},
  \bibinfo{author}{\bibfnamefont{G.}~\bibnamefont{Bollen}},
  \bibinfo{author}{\bibfnamefont{H.-J.} \bibnamefont{Kluge}},
  \bibinfo{author}{\bibfnamefont{T.}~\bibnamefont{Otto}}, \bibnamefont{and}
  \bibinfo{author}{\bibfnamefont{J.}~\bibnamefont{Szerypo}},
  \bibinfo{journal}{Int. J. Mass Spectrom. Ion Process}
  \textbf{\bibinfo{volume}{142}}, \bibinfo{pages}{95} (\bibinfo{year}{1995}),
  \urlprefix\url{https://doi.org/10.1016/0168-1176(95)04146-C.}

\bibitem[{\citenamefont{Kn{\"o}bel et~al.}(2016)\citenamefont{Kn{\"o}bel,
  Diwisch, Geissel, Litvinov, Patyk, Pla{\ss}, Scheidenberger, Sun, Weick,
  Bosch et~al.}}]{Knoebel2016}
\bibinfo{author}{\bibfnamefont{R.}~\bibnamefont{Kn{\"o}bel}},
  \bibinfo{author}{\bibfnamefont{M.}~\bibnamefont{Diwisch}},
  \bibinfo{author}{\bibfnamefont{H.}~\bibnamefont{Geissel}},
  \bibinfo{author}{\bibfnamefont{Y.~A.} \bibnamefont{Litvinov}},
  \bibinfo{author}{\bibfnamefont{Z.}~\bibnamefont{Patyk}},
  \bibinfo{author}{\bibfnamefont{W.~R.} \bibnamefont{Pla{\ss}}},
  \bibinfo{author}{\bibfnamefont{C.}~\bibnamefont{Scheidenberger}},
  \bibinfo{author}{\bibfnamefont{B.}~\bibnamefont{Sun}},
  \bibinfo{author}{\bibfnamefont{H.}~\bibnamefont{Weick}},
  \bibinfo{author}{\bibfnamefont{F.}~\bibnamefont{Bosch}},
  \bibnamefont{et~al.}, \bibinfo{journal}{The European Physical Journal A}
  \textbf{\bibinfo{volume}{52}}, \bibinfo{pages}{138} (\bibinfo{year}{2016}),
  ISSN \bibinfo{issn}{1434-601X},
  \urlprefix\url{https://doi.org/10.1140/epja/i2016-16138-6}.

\bibitem[{\citenamefont{Huang et~al.}(2021)\citenamefont{Huang, Wang, Kondev,
  Audi, and Naimi}}]{Huang2021}
\bibinfo{author}{\bibfnamefont{W.}~\bibnamefont{Huang}},
  \bibinfo{author}{\bibfnamefont{M.}~\bibnamefont{Wang}},
  \bibinfo{author}{\bibfnamefont{F.}~\bibnamefont{Kondev}},
  \bibinfo{author}{\bibfnamefont{G.}~\bibnamefont{Audi}}, \bibnamefont{and}
  \bibinfo{author}{\bibfnamefont{S.}~\bibnamefont{Naimi}},
  \bibinfo{journal}{Chinese Physics C} \textbf{\bibinfo{volume}{45}},
  \bibinfo{pages}{030002} (\bibinfo{year}{2021}),
  \urlprefix\url{https://dx.doi.org/10.1088/1674-1137/abddb0}.

\bibitem[{\citenamefont{Ryssens et~al.}(2022)\citenamefont{Ryssens, Scamps,
  Goriely, and Bender}}]{Ryssens2022}
\bibinfo{author}{\bibfnamefont{W.}~\bibnamefont{Ryssens}},
  \bibinfo{author}{\bibfnamefont{G.}~\bibnamefont{Scamps}},
  \bibinfo{author}{\bibfnamefont{S.}~\bibnamefont{Goriely}}, \bibnamefont{and}
  \bibinfo{author}{\bibfnamefont{M.}~\bibnamefont{Bender}},
  \bibinfo{journal}{The European Physical Journal A}
  \textbf{\bibinfo{volume}{58}}, \bibinfo{pages}{246} (\bibinfo{year}{2022}),
  ISSN \bibinfo{issn}{1434-601X},
  \urlprefix\url{https://doi.org/10.1140/epja/s10050-022-00894-5}.

\bibitem[{\citenamefont{Ryssens et~al.}(2023)\citenamefont{Ryssens, Scamps,
  Goriely, and Bender}}]{Ryssens2023}
\bibinfo{author}{\bibfnamefont{W.}~\bibnamefont{Ryssens}},
  \bibinfo{author}{\bibfnamefont{G.}~\bibnamefont{Scamps}},
  \bibinfo{author}{\bibfnamefont{S.}~\bibnamefont{Goriely}}, \bibnamefont{and}
  \bibinfo{author}{\bibfnamefont{M.}~\bibnamefont{Bender}},
  \bibinfo{journal}{Eur. Phys. J. A} \textbf{\bibinfo{volume}{59}},
  \bibinfo{pages}{96} (\bibinfo{year}{2023}), ISSN \bibinfo{issn}{1434-601X},
  \urlprefix\url{https://doi.org/10.1140/epja/s10050-023-01002-x}.

\bibitem[{\citenamefont{Moore et~al.}(2013)\citenamefont{Moore, Eronen,
  Gorelov, Hakala, Jokinen, Kankainen, Kolhinen, Koponen, Penttil\"a,
  Pohjalainen et~al.}}]{Moore2013}
\bibinfo{author}{\bibfnamefont{I.~D.} \bibnamefont{Moore}},
  \bibinfo{author}{\bibfnamefont{T.}~\bibnamefont{Eronen}},
  \bibinfo{author}{\bibfnamefont{D.}~\bibnamefont{Gorelov}},
  \bibinfo{author}{\bibfnamefont{J.}~\bibnamefont{Hakala}},
  \bibinfo{author}{\bibfnamefont{A.}~\bibnamefont{Jokinen}},
  \bibinfo{author}{\bibfnamefont{A.}~\bibnamefont{Kankainen}},
  \bibinfo{author}{\bibfnamefont{V.}~\bibnamefont{Kolhinen}},
  \bibinfo{author}{\bibfnamefont{J.}~\bibnamefont{Koponen}},
  \bibinfo{author}{\bibfnamefont{H.}~\bibnamefont{Penttil\"a}},
  \bibinfo{author}{\bibfnamefont{I.}~\bibnamefont{Pohjalainen}},
  \bibnamefont{et~al.}, \bibinfo{journal}{Nuc. Inst. and Meth. in Physics
  Research Section B: Beam Interactions with Materials and Atoms}
  \textbf{\bibinfo{volume}{317}}, \bibinfo{pages}{208} (\bibinfo{year}{2013}),
  \urlprefix\url{https://doi.org/10.1016/j.nimb.2013.06.036}.

\bibitem[{\citenamefont{Karvonen et~al.}(2008)\citenamefont{Karvonen, Moore,
  Sonoda, Kessler, Penttilä, Peräjärvi, Ronkanen, and
  Äystö}}]{Karvonen2008}
\bibinfo{author}{\bibfnamefont{P.}~\bibnamefont{Karvonen}},
  \bibinfo{author}{\bibfnamefont{I.~D.} \bibnamefont{Moore}},
  \bibinfo{author}{\bibfnamefont{T.}~\bibnamefont{Sonoda}},
  \bibinfo{author}{\bibfnamefont{T.}~\bibnamefont{Kessler}},
  \bibinfo{author}{\bibfnamefont{H.}~\bibnamefont{Penttilä}},
  \bibinfo{author}{\bibfnamefont{K.}~\bibnamefont{Peräjärvi}},
  \bibinfo{author}{\bibfnamefont{P.}~\bibnamefont{Ronkanen}}, \bibnamefont{and}
  \bibinfo{author}{\bibfnamefont{J.}~\bibnamefont{Äystö}},
  \bibinfo{journal}{Nucl. Instrum. Meth. Phys. Res. B}
  \textbf{\bibinfo{volume}{266}}, \bibinfo{pages}{4794} (\bibinfo{year}{2008}),
  \urlprefix\url{https://doi.org/10.1016/j.nimb.2008.07.022}.

\bibitem[{\citenamefont{Nieminen et~al.}(2001)\citenamefont{Nieminen, Huikari,
  Jokinen, Äystö, Campbell, and Cochrane}}]{Nieminen2001}
\bibinfo{author}{\bibfnamefont{A.}~\bibnamefont{Nieminen}},
  \bibinfo{author}{\bibfnamefont{J.}~\bibnamefont{Huikari}},
  \bibinfo{author}{\bibfnamefont{A.}~\bibnamefont{Jokinen}},
  \bibinfo{author}{\bibfnamefont{J.}~\bibnamefont{Äystö}},
  \bibinfo{author}{\bibfnamefont{P.}~\bibnamefont{Campbell}}, \bibnamefont{and}
  \bibinfo{author}{\bibfnamefont{E.~C.~A.} \bibnamefont{Cochrane}},
  \bibinfo{journal}{Nucl. Instrum. Meth. Phys. Res. A}
  \textbf{\bibinfo{volume}{469}}, \bibinfo{pages}{244} (\bibinfo{year}{2001}),
  \urlprefix\url{https://doi.org/10.1016/S0168-9002(00)00750-6}.

\bibitem[{\citenamefont{Savard et~al.}(1991)\citenamefont{Savard, Becker,
  Bollen, Kluge, Moore, Otto, Schweikhard, Stolzenberg, and
  Wiess}}]{Savard1991}
\bibinfo{author}{\bibfnamefont{G.}~\bibnamefont{Savard}},
  \bibinfo{author}{\bibfnamefont{S.}~\bibnamefont{Becker}},
  \bibinfo{author}{\bibfnamefont{G.}~\bibnamefont{Bollen}},
  \bibinfo{author}{\bibfnamefont{H.-J.} \bibnamefont{Kluge}},
  \bibinfo{author}{\bibfnamefont{R.~B.} \bibnamefont{Moore}},
  \bibinfo{author}{\bibfnamefont{T.}~\bibnamefont{Otto}},
  \bibinfo{author}{\bibfnamefont{L.}~\bibnamefont{Schweikhard}},
  \bibinfo{author}{\bibfnamefont{H.}~\bibnamefont{Stolzenberg}},
  \bibnamefont{and} \bibinfo{author}{\bibfnamefont{U.}~\bibnamefont{Wiess}},
  \bibinfo{journal}{Phys. Lett. A} \textbf{\bibinfo{volume}{158}},
  \bibinfo{pages}{247} (\bibinfo{year}{1991}),
  \urlprefix\url{https://doi.org/10.1016/0375-9601(91)91008-2}.

\bibitem[{\citenamefont{Kondev et~al.}(2021)\citenamefont{Kondev, Wang, Huang,
  Naimi, and Audi}}]{NUBASE20}
\bibinfo{author}{\bibfnamefont{F.}~\bibnamefont{Kondev}},
  \bibinfo{author}{\bibfnamefont{M.}~\bibnamefont{Wang}},
  \bibinfo{author}{\bibfnamefont{W.}~\bibnamefont{Huang}},
  \bibinfo{author}{\bibfnamefont{S.}~\bibnamefont{Naimi}}, \bibnamefont{and}
  \bibinfo{author}{\bibfnamefont{G.}~\bibnamefont{Audi}},
  \bibinfo{journal}{Chinese Physics C} \textbf{\bibinfo{volume}{45}},
  \bibinfo{pages}{030001} (\bibinfo{year}{2021}),
  \urlprefix\url{https://doi.org/10.1088/1674-1137/abddae}.

\bibitem[{\citenamefont{Vilén et~al.}(2020)\citenamefont{Vilén, Canete,
  Cheal, Giatzoglou, {de Groote}, {de Roubin}, Eronen, Geldhof, Jokinen,
  Kankainen et~al.}}]{Vilen2020}
\bibinfo{author}{\bibfnamefont{M.}~\bibnamefont{Vilén}},
  \bibinfo{author}{\bibfnamefont{L.}~\bibnamefont{Canete}},
  \bibinfo{author}{\bibfnamefont{B.}~\bibnamefont{Cheal}},
  \bibinfo{author}{\bibfnamefont{A.}~\bibnamefont{Giatzoglou}},
  \bibinfo{author}{\bibfnamefont{R.}~\bibnamefont{{de Groote}}},
  \bibinfo{author}{\bibfnamefont{A.}~\bibnamefont{{de Roubin}}},
  \bibinfo{author}{\bibfnamefont{T.}~\bibnamefont{Eronen}},
  \bibinfo{author}{\bibfnamefont{S.}~\bibnamefont{Geldhof}},
  \bibinfo{author}{\bibfnamefont{A.}~\bibnamefont{Jokinen}},
  \bibinfo{author}{\bibfnamefont{A.}~\bibnamefont{Kankainen}},
  \bibnamefont{et~al.}, \bibinfo{journal}{Nuclear Instruments and Methods in
  Physics Research Section B: Beam Interactions with Materials and Atoms}
  \textbf{\bibinfo{volume}{463}}, \bibinfo{pages}{382} (\bibinfo{year}{2020}),
  ISSN \bibinfo{issn}{0168-583X},
  \urlprefix\url{https://www.sciencedirect.com/science/article/pii/S0168583X19302344}.

\bibitem[{\citenamefont{Nesterenko et~al.}(2018)\citenamefont{Nesterenko,
  Eronen, Kankainen, Canete, Jokinen, Moore, Penttilä, Rinta-Antila,
  de~Roubin, and Vilen}}]{Nesterenko2018}
\bibinfo{author}{\bibfnamefont{D.~A.} \bibnamefont{Nesterenko}},
  \bibinfo{author}{\bibfnamefont{T.}~\bibnamefont{Eronen}},
  \bibinfo{author}{\bibfnamefont{A.}~\bibnamefont{Kankainen}},
  \bibinfo{author}{\bibfnamefont{L.}~\bibnamefont{Canete}},
  \bibinfo{author}{\bibfnamefont{A.}~\bibnamefont{Jokinen}},
  \bibinfo{author}{\bibfnamefont{I.~D.} \bibnamefont{Moore}},
  \bibinfo{author}{\bibfnamefont{H.}~\bibnamefont{Penttilä}},
  \bibinfo{author}{\bibfnamefont{S.}~\bibnamefont{Rinta-Antila}},
  \bibinfo{author}{\bibfnamefont{A.}~\bibnamefont{de~Roubin}},
  \bibnamefont{and} \bibinfo{author}{\bibfnamefont{M.}~\bibnamefont{Vilen}},
  \bibinfo{journal}{Eur. Phys. J. A.} \textbf{\bibinfo{volume}{54}},
  \bibinfo{pages}{154} (\bibinfo{year}{2018}),
  \urlprefix\url{https://doi.org/10.1140/epja/i2018-12589-y}.

\bibitem[{\citenamefont{Nesterenko et~al.}(2021)\citenamefont{Nesterenko,
  Eronen, Ge, Kankainen, and Vilen}}]{Nesterenko2021}
\bibinfo{author}{\bibfnamefont{D.}~\bibnamefont{Nesterenko}},
  \bibinfo{author}{\bibfnamefont{T.}~\bibnamefont{Eronen}},
  \bibinfo{author}{\bibfnamefont{Z.}~\bibnamefont{Ge}},
  \bibinfo{author}{\bibfnamefont{A.}~\bibnamefont{Kankainen}},
  \bibnamefont{and} \bibinfo{author}{\bibfnamefont{M.}~\bibnamefont{Vilen}},
  \bibinfo{journal}{Eur. Phys. Jour. A} \textbf{\bibinfo{volume}{57}},
  \bibinfo{pages}{302} (\bibinfo{year}{2021}),
  \urlprefix\url{https://doi.org/10.1140/epja/s10050-021-00608-3}.

\bibitem[{\citenamefont{Gräff et~al.}(1980)\citenamefont{Gräff, Kalinowsky,
  and Traut}}]{Graff1980}
\bibinfo{author}{\bibfnamefont{G.}~\bibnamefont{Gräff}},
  \bibinfo{author}{\bibfnamefont{H.}~\bibnamefont{Kalinowsky}},
  \bibnamefont{and} \bibinfo{author}{\bibfnamefont{J.}~\bibnamefont{Traut}},
  \bibinfo{journal}{Z Physik A} \textbf{\bibinfo{volume}{297}},
  \bibinfo{pages}{35} (\bibinfo{year}{1980}),
  \urlprefix\url{https://doi.org/10.1007/BF01414243}.

\bibitem[{\citenamefont{Kretzschmar}(2007)}]{Kretzschmar2007}
\bibinfo{author}{\bibfnamefont{M.}~\bibnamefont{Kretzschmar}},
  \bibinfo{journal}{Int. J. Mass Spectrom.} \textbf{\bibinfo{volume}{264}},
  \bibinfo{pages}{122} (\bibinfo{year}{2007}),
  \urlprefix\url{https://doi.org/10.1016/j.ijms.2007.04.002}.

\bibitem[{\citenamefont{George et~al.}(2007)\citenamefont{George, Blaum,
  Herfurth, Herlert, Kretzschmar, Nagy, Schwarz, Schweikhard, and
  Yazidjian}}]{George2007}
\bibinfo{author}{\bibfnamefont{S.}~\bibnamefont{George}},
  \bibinfo{author}{\bibfnamefont{K.}~\bibnamefont{Blaum}},
  \bibinfo{author}{\bibfnamefont{F.}~\bibnamefont{Herfurth}},
  \bibinfo{author}{\bibfnamefont{A.}~\bibnamefont{Herlert}},
  \bibinfo{author}{\bibfnamefont{M.}~\bibnamefont{Kretzschmar}},
  \bibinfo{author}{\bibfnamefont{S.}~\bibnamefont{Nagy}},
  \bibinfo{author}{\bibfnamefont{S.}~\bibnamefont{Schwarz}},
  \bibinfo{author}{\bibfnamefont{L.}~\bibnamefont{Schweikhard}},
  \bibnamefont{and}
  \bibinfo{author}{\bibfnamefont{C.}~\bibnamefont{Yazidjian}},
  \bibinfo{journal}{Int. J. Mass Spectrom.} \textbf{\bibinfo{volume}{264}},
  \bibinfo{pages}{110} (\bibinfo{year}{2007}),
  \urlprefix\url{https://doi.org/10.1016/j.ijms.2007.04.003}.

\bibitem[{\citenamefont{Kellerbauer et~al.}(2003)\citenamefont{Kellerbauer,
  Blaum, Bollen, Herfurth, Kluge, Kuckein, Sauvan, Scheidenberger, and
  Schweikhard}}]{Kellerbauer2003}
\bibinfo{author}{\bibfnamefont{A.}~\bibnamefont{Kellerbauer}},
  \bibinfo{author}{\bibfnamefont{K.}~\bibnamefont{Blaum}},
  \bibinfo{author}{\bibfnamefont{G.}~\bibnamefont{Bollen}},
  \bibinfo{author}{\bibfnamefont{F.}~\bibnamefont{Herfurth}},
  \bibinfo{author}{\bibfnamefont{H.-J.} \bibnamefont{Kluge}},
  \bibinfo{author}{\bibfnamefont{M.}~\bibnamefont{Kuckein}},
  \bibinfo{author}{\bibfnamefont{E.}~\bibnamefont{Sauvan}},
  \bibinfo{author}{\bibfnamefont{C.}~\bibnamefont{Scheidenberger}},
  \bibnamefont{and}
  \bibinfo{author}{\bibfnamefont{L.}~\bibnamefont{Schweikhard}},
  \bibinfo{journal}{Eur. Phys. J. D} \textbf{\bibinfo{volume}{22}},
  \bibinfo{pages}{53} (\bibinfo{year}{2003}),
  \urlprefix\url{https://doi.org/10.1140/epjd/e2002-00222-0}.

\bibitem[{\citenamefont{Wang et~al.}(2021)\citenamefont{Wang, Huang, Kondev,
  Audi, and Naimi}}]{AME20}
\bibinfo{author}{\bibfnamefont{M.}~\bibnamefont{Wang}},
  \bibinfo{author}{\bibfnamefont{W.}~\bibnamefont{Huang}},
  \bibinfo{author}{\bibfnamefont{F.}~\bibnamefont{Kondev}},
  \bibinfo{author}{\bibfnamefont{G.}~\bibnamefont{Audi}}, \bibnamefont{and}
  \bibinfo{author}{\bibfnamefont{S.}~\bibnamefont{Naimi}},
  \bibinfo{journal}{Chinese Physics C} \textbf{\bibinfo{volume}{45}},
  \bibinfo{pages}{030003} (\bibinfo{year}{2021}),
  \urlprefix\url{https://doi.org/10.1088/1674-1137/abddaf}.

\bibitem[{\citenamefont{Rissanen et~al.}(2011)\citenamefont{Rissanen, Kurpeta,
  Plochocki, Elomaa, Eronen, Hakala, Jokinen, Kankainen, Karvonen, Moore
  et~al.}}]{Rissanen2011}
\bibinfo{author}{\bibfnamefont{J.}~\bibnamefont{Rissanen}},
  \bibinfo{author}{\bibfnamefont{J.}~\bibnamefont{Kurpeta}},
  \bibinfo{author}{\bibfnamefont{A.}~\bibnamefont{Plochocki}},
  \bibinfo{author}{\bibfnamefont{V.~V.} \bibnamefont{Elomaa}},
  \bibinfo{author}{\bibfnamefont{T.}~\bibnamefont{Eronen}},
  \bibinfo{author}{\bibfnamefont{J.}~\bibnamefont{Hakala}},
  \bibinfo{author}{\bibfnamefont{A.}~\bibnamefont{Jokinen}},
  \bibinfo{author}{\bibfnamefont{A.}~\bibnamefont{Kankainen}},
  \bibinfo{author}{\bibfnamefont{P.}~\bibnamefont{Karvonen}},
  \bibinfo{author}{\bibfnamefont{I.~D.} \bibnamefont{Moore}},
  \bibnamefont{et~al.}, \bibinfo{journal}{Eur. Phys. J. A}
  \textbf{\bibinfo{volume}{47}}, \bibinfo{pages}{97} (\bibinfo{year}{2011}),
  \urlprefix\url{https://doi.org/10.1140/epja/i2011-11097-0}.

\bibitem[{\citenamefont{Audi et~al.}(2003)\citenamefont{Audi, Bersillon,
  Blachot, and Wapstra}}]{NUBASE03}
\bibinfo{author}{\bibfnamefont{G.}~\bibnamefont{Audi}},
  \bibinfo{author}{\bibfnamefont{O.}~\bibnamefont{Bersillon}},
  \bibinfo{author}{\bibfnamefont{J.}~\bibnamefont{Blachot}}, \bibnamefont{and}
  \bibinfo{author}{\bibfnamefont{A.}~\bibnamefont{Wapstra}},
  \bibinfo{journal}{Nucl. Phys. A} \textbf{\bibinfo{volume}{729}},
  \bibinfo{pages}{3} (\bibinfo{year}{2003}), ISSN \bibinfo{issn}{0375-9474},
  \bibinfo{note}{the 2003 NUBASE and Atomic Mass Evaluations},
  \urlprefix\url{https://www.sciencedirect.com/science/article/pii/S0375947403018074}.

\bibitem[{\citenamefont{Audi et~al.}(2012)\citenamefont{Audi, Kondev, Wang,
  Pfeiffer, Sun, Blachot, and MacCormick}}]{NUBASE12}
\bibinfo{author}{\bibfnamefont{G.}~\bibnamefont{Audi}},
  \bibinfo{author}{\bibfnamefont{F.~G.} \bibnamefont{Kondev}},
  \bibinfo{author}{\bibfnamefont{M.}~\bibnamefont{Wang}},
  \bibinfo{author}{\bibfnamefont{B.}~\bibnamefont{Pfeiffer}},
  \bibinfo{author}{\bibfnamefont{X.}~\bibnamefont{Sun}},
  \bibinfo{author}{\bibfnamefont{J.}~\bibnamefont{Blachot}}, \bibnamefont{and}
  \bibinfo{author}{\bibfnamefont{M.}~\bibnamefont{MacCormick}},
  \bibinfo{journal}{Chin. Phys. C} \textbf{\bibinfo{volume}{36}},
  \bibinfo{pages}{1157} (\bibinfo{year}{2012}),
  \urlprefix\url{https://dx.doi.org/10.1088/1674-1137/36/12/001}.

\bibitem[{\citenamefont{Audi et~al.}(2017)\citenamefont{Audi, Kondev, Wang,
  Huang, and Naimi}}]{NUBASE16}
\bibinfo{author}{\bibfnamefont{G.}~\bibnamefont{Audi}},
  \bibinfo{author}{\bibfnamefont{F.~G.} \bibnamefont{Kondev}},
  \bibinfo{author}{\bibfnamefont{M.}~\bibnamefont{Wang}},
  \bibinfo{author}{\bibfnamefont{W.~J.} \bibnamefont{Huang}}, \bibnamefont{and}
  \bibinfo{author}{\bibfnamefont{S.}~\bibnamefont{Naimi}},
  \bibinfo{journal}{Chin. Phys. C} \textbf{\bibinfo{volume}{41}},
  \bibinfo{pages}{030001} (\bibinfo{year}{2017}),
  \urlprefix\url{https://dx.doi.org/10.1088/1674-1137/41/3/030001}.

\bibitem[{\citenamefont{{J. Kurpeta} et~al.}(1998)\citenamefont{{J. Kurpeta},
  {G. Lhersonneau}, {J.C. Wang}, {P. Dendooven}, {A. Honkanen}, {M. Huhta}, {M.
  Oinonen}, {H. Penttil\"a}, {K. Per\"aj\"arvi}, {J.R. Persson}
  et~al.}}]{Kurpeta1998}
\bibinfo{author}{\bibnamefont{{J. Kurpeta}}}, \bibinfo{author}{\bibnamefont{{G.
  Lhersonneau}}}, \bibinfo{author}{\bibnamefont{{J.C. Wang}}},
  \bibinfo{author}{\bibnamefont{{P. Dendooven}}},
  \bibinfo{author}{\bibnamefont{{A. Honkanen}}},
  \bibinfo{author}{\bibnamefont{{M. Huhta}}}, \bibinfo{author}{\bibnamefont{{M.
  Oinonen}}}, \bibinfo{author}{\bibnamefont{{H. Penttil\"a}}},
  \bibinfo{author}{\bibnamefont{{K. Per\"aj\"arvi}}},
  \bibinfo{author}{\bibnamefont{{J.R. Persson}}}, \bibnamefont{et~al.},
  \bibinfo{journal}{Eur. Phys. J. A} \textbf{\bibinfo{volume}{2}},
  \bibinfo{pages}{241} (\bibinfo{year}{1998}),
  \urlprefix\url{https://doi.org/10.1007/s100500050114}.

\bibitem[{\citenamefont{Kurpeta et~al.}(2007)\citenamefont{Kurpeta, Urban,
  Droste, P{\l}ochocki, Rohozi{\'{n}}ski, Rzaca-Urban, Morek, Pr{\'o}chniak,
  Starosta, {\"A}yst{\"o} et~al.}}]{Kurpeta2007}
\bibinfo{author}{\bibfnamefont{J.}~\bibnamefont{Kurpeta}},
  \bibinfo{author}{\bibfnamefont{W.}~\bibnamefont{Urban}},
  \bibinfo{author}{\bibfnamefont{C.}~\bibnamefont{Droste}},
  \bibinfo{author}{\bibfnamefont{A.}~\bibnamefont{P{\l}ochocki}},
  \bibinfo{author}{\bibfnamefont{S.~G.} \bibnamefont{Rohozi{\'{n}}ski}},
  \bibinfo{author}{\bibfnamefont{T.}~\bibnamefont{Rzaca-Urban}},
  \bibinfo{author}{\bibfnamefont{T.}~\bibnamefont{Morek}},
  \bibinfo{author}{\bibfnamefont{L.}~\bibnamefont{Pr{\'o}chniak}},
  \bibinfo{author}{\bibfnamefont{K.}~\bibnamefont{Starosta}},
  \bibinfo{author}{\bibfnamefont{J.}~\bibnamefont{{\"A}yst{\"o}}},
  \bibnamefont{et~al.}, \bibinfo{journal}{The European Physical Journal A}
  \textbf{\bibinfo{volume}{33}}, \bibinfo{pages}{307} (\bibinfo{year}{2007}),
  ISSN \bibinfo{issn}{1434-601X},
  \urlprefix\url{https://doi.org/10.1140/epja/i2006-10464-2}.

\bibitem[{\citenamefont{Blachot}(2010)}]{Blachot2010}
\bibinfo{author}{\bibfnamefont{J.}~\bibnamefont{Blachot}},
  \bibinfo{journal}{Nucl. Data Sheets} \textbf{\bibinfo{volume}{111}},
  \bibinfo{pages}{1471} (\bibinfo{year}{2010}), ISSN \bibinfo{issn}{0090-3752},
  \urlprefix\url{https://www.sciencedirect.com/science/article/pii/S0090375210000529}.

\bibitem[{\citenamefont{Kratz and Pfeiffer}()}]{Kratz2000}
\bibinfo{author}{\bibfnamefont{K.-L.} \bibnamefont{Kratz}} \bibnamefont{and}
  \bibinfo{author}{\bibfnamefont{B.}~\bibnamefont{Pfeiffer}},
  \bibinfo{note}{private communication to G. Audi, June 2000}.

\bibitem[{\citenamefont{Kurpeta et~al.}(2010)\citenamefont{Kurpeta, Rissanen,
  P\l{}ochocki, Urban, Elomaa, Eronen, Hakala, Jokinen, Kankainen, Karvonen
  et~al.}}]{Kurpeta2010}
\bibinfo{author}{\bibfnamefont{J.}~\bibnamefont{Kurpeta}},
  \bibinfo{author}{\bibfnamefont{J.}~\bibnamefont{Rissanen}},
  \bibinfo{author}{\bibfnamefont{A.}~\bibnamefont{P\l{}ochocki}},
  \bibinfo{author}{\bibfnamefont{W.}~\bibnamefont{Urban}},
  \bibinfo{author}{\bibfnamefont{V.-V.} \bibnamefont{Elomaa}},
  \bibinfo{author}{\bibfnamefont{T.}~\bibnamefont{Eronen}},
  \bibinfo{author}{\bibfnamefont{J.}~\bibnamefont{Hakala}},
  \bibinfo{author}{\bibfnamefont{A.}~\bibnamefont{Jokinen}},
  \bibinfo{author}{\bibfnamefont{A.}~\bibnamefont{Kankainen}},
  \bibinfo{author}{\bibfnamefont{P.}~\bibnamefont{Karvonen}},
  \bibnamefont{et~al.}, \bibinfo{journal}{Phys. Rev. C}
  \textbf{\bibinfo{volume}{82}}, \bibinfo{pages}{064318}
  (\bibinfo{year}{2010}),
  \urlprefix\url{https://link.aps.org/doi/10.1103/PhysRevC.82.064318}.

\bibitem[{\citenamefont{Matos}(2004)}]{Matos2004}
\bibinfo{author}{\bibfnamefont{M.}~\bibnamefont{Matos}}, Ph.D. thesis,
  \bibinfo{school}{Justus-Liebig-Universit\"at Giessen} (\bibinfo{year}{2004}),
  \urlprefix\url{http://dx.doi.org/10.22029/jlupub-9512}.

\bibitem[{\citenamefont{Goriely et~al.}(2016)\citenamefont{Goriely, Chamel, and
  Pearson}}]{goriely2016}
\bibinfo{author}{\bibfnamefont{S.}~\bibnamefont{Goriely}},
  \bibinfo{author}{\bibfnamefont{N.}~\bibnamefont{Chamel}}, \bibnamefont{and}
  \bibinfo{author}{\bibfnamefont{J.~M.} \bibnamefont{Pearson}},
  \bibinfo{journal}{Physical Review C} \textbf{\bibinfo{volume}{93}},
  \bibinfo{pages}{034337} (\bibinfo{year}{2016}), ISSN
  \bibinfo{issn}{2469-9985, 2469-9993},
  \urlprefix\url{https://link.aps.org/doi/10.1103/PhysRevC.93.034337}.

\bibitem[{\citenamefont{Möller et~al.}(2016)\citenamefont{Möller, Sierk,
  Ichikawa, and Sagawa}}]{Moller2016}
\bibinfo{author}{\bibfnamefont{P.}~\bibnamefont{Möller}},
  \bibinfo{author}{\bibfnamefont{A.}~\bibnamefont{Sierk}},
  \bibinfo{author}{\bibfnamefont{T.}~\bibnamefont{Ichikawa}}, \bibnamefont{and}
  \bibinfo{author}{\bibfnamefont{H.}~\bibnamefont{Sagawa}},
  \bibinfo{journal}{Atomic Data and Nuclear Data Tables}
  \textbf{\bibinfo{volume}{109-110}}, \bibinfo{pages}{1}
  (\bibinfo{year}{2016}), ISSN \bibinfo{issn}{0092-640X},
  \urlprefix\url{https://www.sciencedirect.com/science/article/pii/S0092640X1600005X}.

\bibitem[{\citenamefont{Duflo and Zuker}(1995)}]{duflo_1995}
\bibinfo{author}{\bibfnamefont{J.}~\bibnamefont{Duflo}} \bibnamefont{and}
  \bibinfo{author}{\bibfnamefont{A.}~\bibnamefont{Zuker}},
  \bibinfo{journal}{Physical Review C} \textbf{\bibinfo{volume}{52}},
  \bibinfo{pages}{R23} (\bibinfo{year}{1995}), ISSN \bibinfo{issn}{0556-2813,
  1089-490X}, \urlprefix\url{https://link.aps.org/doi/10.1103/PhysRevC.52.R23}.

\bibitem[{\citenamefont{Niu and Liang}(2022)}]{niu_2022}
\bibinfo{author}{\bibfnamefont{Z.~M.} \bibnamefont{Niu}} \bibnamefont{and}
  \bibinfo{author}{\bibfnamefont{H.~Z.} \bibnamefont{Liang}},
  \bibinfo{journal}{Physical Review C} \textbf{\bibinfo{volume}{106}},
  \bibinfo{pages}{L021303} (\bibinfo{year}{2022}),
  \urlprefix\url{https://link.aps.org/doi/10.1103/PhysRevC.106.L021303}.

\bibitem[{\citenamefont{Ryssens}(2016)}]{Ryssens2016}
\bibinfo{author}{\bibfnamefont{W.}~\bibnamefont{Ryssens}}, Ph.D. thesis,
  \bibinfo{school}{Université Libre de Bruxelles}, \bibinfo{address}{Brussels}
  (\bibinfo{year}{2016}),
  \urlprefix\url{https://inspirehep.net/files/3011a13e26b3c2dbf7429a4020f855bd}.

\bibitem[{\citenamefont{Xu et~al.}(2002)\citenamefont{Xu, Walker, and
  Wyss}}]{Xu_2002}
\bibinfo{author}{\bibfnamefont{F.~R.} \bibnamefont{Xu}},
  \bibinfo{author}{\bibfnamefont{P.~M.} \bibnamefont{Walker}},
  \bibnamefont{and} \bibinfo{author}{\bibfnamefont{R.}~\bibnamefont{Wyss}},
  \bibinfo{journal}{Phys. Rev. C} \textbf{\bibinfo{volume}{65}},
  \bibinfo{pages}{021303} (\bibinfo{year}{2002}),
  \urlprefix\url{https://link.aps.org/doi/10.1103/PhysRevC.65.021303}.

\bibitem[{\citenamefont{Bally and Bender}(2021)}]{Bally21a}
\bibinfo{author}{\bibfnamefont{B.}~\bibnamefont{Bally}} \bibnamefont{and}
  \bibinfo{author}{\bibfnamefont{M.}~\bibnamefont{Bender}},
  \bibinfo{journal}{Phys. Rev. C} \textbf{\bibinfo{volume}{103}},
  \bibinfo{pages}{024315} (\bibinfo{year}{2021}),
  \urlprefix\url{https://link.aps.org/doi/10.1103/PhysRevC.103.024315}.

\end{thebibliography}

\end{document}